\documentclass[aip,reprint]{revtex4-1}
\usepackage{graphicx}
\usepackage{gensymb}
\usepackage{amssymb}
\usepackage{amsmath}
\usepackage{xcolor}
\usepackage[normalem]{ulem}

\begin{document}

\title{INFRA-ICE: an ultra-high vacuum experimental station for laboratory astrochemistry}

\author{Gonzalo Santoro}
\email[Authors to whom correspondence should be addressed:\\]{gonzalo.santoro@icmm.csic.es; gago@icmm.csic.es}
\affiliation{Instituto de Ciencia de Materiales de Madrid (ICMM, CSIC). Materials Science Factory. Structure of Nanoscopic Systems Group. c/ Sor Juana In\'es de la Cruz 3, E-28049 Cantoblanco, Madrid, Spain.}

\author{Jes\'us. M. Sobrado}
\affiliation{Centro de Astrobiolog\'ia (CAB, INTA-CSIC). Crta. de Torrej\'on a Ajalvir km4, E-28850, Torrejón de Ardoz, Madrid, Spain.}

\author{Guillermo Tajuelo-Castilla}
\affiliation{Instituto de Ciencia de Materiales de Madrid (ICMM, CSIC). Materials Science Factory. Structure of Nanoscopic Systems Group. c/ Sor Juana In\'es de la Cruz 3, E-28049 Cantoblanco, Madrid, Spain.}

\author{Mario Accolla}
\affiliation{Instituto de Ciencia de Materiales de Madrid (ICMM, CSIC). Materials Science Factory. Structure of Nanoscopic Systems Group. c/ Sor Juana In\'es de la Cruz 3, E-28049 Cantoblanco, Madrid, Spain.}

\author{Lidia Mart\'inez}
\affiliation{Instituto de Ciencia de Materiales de Madrid (ICMM, CSIC). Materials Science Factory. Structure of Nanoscopic Systems Group. c/ Sor Juana In\'es de la Cruz 3, E-28049 Cantoblanco, Madrid, Spain.}

\author{Jon Azpeitia}
\affiliation{Instituto de Ciencia de Materiales de Madrid (ICMM, CSIC). Materials Science Factory. Structure of Nanoscopic Systems Group. c/ Sor Juana In\'es de la Cruz 3, E-28049 Cantoblanco, Madrid, Spain.}

\author{Koen Lauwaet}
\affiliation{IMDEA Nanociencia. Ciudad Universitaria de Cantoblanco, E-28049 Cantoblanco, Madrid, Spain}

\author{Jos\'e Cernicharo}
\affiliation{Instituto de F\'isica Fundamental (IFF, CSIC). Group of Molecular Astrophysics. c/ Serrano 123, 28006 Madrid, Spain.}

\author{Gary J. Ellis}
\affiliation{Instituto de Ciencia y Tecnolog\'ia de Polímeros (ICTP, CSIC). c/ Juan de la Cierva 3, E-28006 Madrid, Spain.}

\author{Jos\'e \'Angel Mart\'in-Gago}
\email[Authors to whom correspondence should be addressed:\\]{gonzalo.santoro@icmm.csic.es; gago@icmm.csic.es}
\affiliation{Instituto de Ciencia de Materiales de Madrid (ICMM, CSIC). Materials Science Factory. Structure of Nanoscopic Systems Group. c/ Sor Juana In\'es de la Cruz 3, E-28049 Cantoblanco, Madrid, Spain.}

\date{\today}

\begin{abstract}
Laboratory astrochemistry aims at simulating in the laboratory some of the chemical and physical processes that operate in different regions of the Universe. Amongst the diverse astrochemical problems that can be addressed in the laboratory, the evolution of cosmic dust grains in the different regions of the interstellar medium (ISM)  and its role in the formation of new chemical species through catalytic processes present significant interest. In particular, in the dark clouds of the ISM dust grains are coated by icy mantles and it is thought that the ice-dust interaction plays a crucial role in the development of the chemical complexity observed in space. Here, we present a new ultra-high vacuum experimental station devoted to simulate the complex conditions of the coldest regions of the ISM. The INFRA-ICE machine can be operated as a standing alone setup or incorporated in a larger experimental station called \textit{Stardust}, which is dedicated to simulate the formation of cosmic dust in evolved stars. As such, INFRA-ICE  expands the capabilities of \textit{Stardust} allowing the simulation of the complete journey of cosmic dust in space, from its formation in asymptotic giant branch stars (AGBs) to its processing and interaction with icy mantles in molecular clouds. To demonstrate some of the capabilities of INFRA-ICE, we present selected results on the UV photochemistry of undecane (C$_{11}$H$_{24}$) at 14 K. Aliphatics are part of the carbonaceous cosmic dust and, recently, aliphatics and short \textit{n}-alkanes have been detected \textit{in-situ} in the comet 67P/Churyumov-Gerasimenko.

\end{abstract}
\maketitle

\section{Introduction}

Laboratory astrophysics and astrochemistry constitutes a very powerful tool for investigating the fundamental physical and chemical processes governing the evolution of matter in space. By simulating in the laboratory the conditions of different regions of the Universe, it is possible not only to test hypotheses derived from astronomical observations and models, but also to provide the astronomers with plausible chemical and physical mechanisms that may operate in space, which would help in the correct interpretation of the results derived from observations and physico-chemical modelling. Thus, laboratory astrophysics arises from the interplay between astronomers, physicists and chemists to synergistically address the chemical evolution of matter in the Universe.

Laboratory astrochemistry encompasses diverse topics related to the chemistry of different regions of the Universe comprising among others, the gas-phase chemistry of species relevant to the chemical evolution in space \cite{canosa97, antinolo16, potapov17,tanarro18,cernicharo19}, the simulation of planetary atmospheres\cite{sobrado14,sobrado15}, the spectroscopic characterization of radicals and ions \cite{romanini99,brechignac99,joblin02,biennier03,kaiser05a,useli10,asvany10,campbell17,domenech18,fernandez19} and the simulation of the circumstellar envelopes (CSEs) of asymptotic giant branch stars (AGBs) \cite{jager09,contreras13,martinez20,santoro20} as well as of the different interstellar environments. \cite{roser01,mennella06,oba09,palumbo10,munozcaro10,linnartz15,fulvio17,hudson17,potapov19a,oba19}.

Among the diverse open questions concerning the development of molecular complexity in space, the role of cosmic dust grains deserves particular attention. Cosmic dust is mainly formed in the CSEs of AGBs and is subsequently ejected into the interstellar medium (ISM) and coated by icy mantles in the molecular clouds by the condensation of gas-phase molecules \cite{boogert15}. During this long journey from the parent star to the ISM, cosmic dust cools down from 1000-1500 K in the dust formation region of the AGBs to around 10 K in the coldest regions of the ISM \cite{pascoli00,contreras13}. In addition, dust grains are subjected to energetic processing, starting in the outer layers of the star where galactic UV photons penetrate the CSE initiating a very rich photochemistry, and continuing in the molecular clouds, where the volume density of molecules is 10$^{2}$-10${^6}$ mol cm$^{-3}$. From an astrochemical point of view, cosmic dust is believed to actively participate in the synthesis of molecules in space by catalyzing chemical reactions on its surface\cite{vidali13,williams16,wakelam17}. However, much is still lacking in the understanding of the exact nature of the catalytic role of dust grains as well as their interactions with icy mantles in dense molecular clouds.

Chemical reactions in the ISM are induced by the processing of matter by cosmic rays and UV photons, both in the gas and in the solid phase \cite{allamandola99}. Cosmic rays (protons, nuclei of heavy atoms, alpha particles, electrons) are ubiquitous throughout the ISM and, in the obscured regions of interstellar clouds where UV photons cannot penetrate, cosmic rays ionize the gaseous species being the gas-phase chemistry dominated by ion-neutral reactions. On the other hand, UV photons reach the edge of interstellar clouds and control the chemistry by destroying most chemical species in these regions and producing new molecules. 

In the obscure, cold and dense regions of the ISM most molecules are condensed into the surfaces of dust grains in the form of molecular ices and a very rich chemistry is initiated by the UV field leading to the formation of complex organic molecules. Moreover, the impact of ions on the icy mantles leads to the physico-chemical modification of the ices, which can also promote the formation of new species\cite{strazzulla01}. The new molecules synthesized as a consequence of the UV and/or ion processing, can be incorporated into to the gas-phase by ion sputtering of the solid material\cite{johnson98} as well as by the increase in temperature of the of the dust/ice system as a result of, e.g., the explosion of a nearby supernova or the increase in temperature in a protoplanetary disk.

The UV photochemistry of molecular ices has been extensively studied in the laboratory \cite{oberg16}. For instance, the formation of aminoacids has been observed from the simplest, glycine, to more complex aminoacids such as serine and aspartic acid through the UV irradiation of molecular ice mixtures resembling the composition of interstellar ices \cite{munozcaro02}. More recently, the central molecular subunit of RNA, ribose has been synthesized in the laboratory by UV irradiation of precometary ices analogs \cite{meinert16}. On the other hand, cosmic rays are highly ionizing radiation and the ions produced by the impact of cosmic rays can, as abovementioned, interact with dust grains and icy mantles producing sputtering of the material and promoting chemical reactions\cite{strazzulla01}. In addition, the interaction of cosmic rays and solid matter induces a cascade of secondary electrons that can participate in ice-grain chemistry (e.g., high energy electron irradiation of silane at low temperature has been shown to promote the formation of Si$_{n}$H$_{2n+2}$ molecules, therefore inducing the growth of polysilanes \cite{tarczay16}, whereas the exposure of a CO$_2$:CH$_4$:NH$_3$ ice mixture to low energy electrons has proved the synthesis of glycine\cite{esmaili18}). Finally, surface etching of SiC grains by exposure to atomic hydrogen, has been shown to generate large polycyclic aromatic hydrocarbons (PAHs)\cite{merino14}.

To address the catalytic activity of cosmic dust, laboratory  studies are mandatory. In this sense, experimental setups specially devoted to investigate the chemical reactions at low temperature on the surface of cosmic dust analogs have been developed\cite{fraser04,ioppolo13,potapov19a}. Recently, the catalytic effect of carbonaceous cosmic dust analogs in the reaction of ammonia and carbon dioxide has been demonstrated \cite{potapov19b} and the utmost importance of the surface of dust grains on the chemistry when the grains are coated by few monolayers of ice has been emphasized \cite{potapov20}. This reflects the need for precise and controlled laboratory simulations of the chemistry taking place in the coldest regions of the ISM by the interaction of dust and molecular ices when subjected to  processing by UV radiation, ions and/or electrons. Realistic cosmic dust analogs, i.e., synthesized at conditions resembling those of the dust formation regions of AGBs, and processing of their covering ices, are crucial in mimicking the chemistry of the ISM.

Here we present the INFRA-ICE experimental station that is devoted to investigate the interaction of cosmic dust and molecular ices under conditions resembling those of the coldest regions of ISM, where dust grains are covered with icy mantles. This experimental station can be incorporated as a new module of the \textit{Stardust} machine \cite{martinez18, martinez20, santoro20}, which is dedicated to simulate in the laboratory the complex conditions that lead to the formation of cosmic dust in the CSEs of AGBs, or it can be operated independently as an autonomous experimental station. When incorporated into \textit{Stardust}, the INFRA-ICE module expands the capabilities of \textit{Stardust} allowing the study of dust-ice interactions (using dust analogs produced \textit{in situ}) and the subsequent chemistry that takes place during the processing of these materials in the dark clouds of the ISM. In order to demonstrate some of the capabilities of INFRA-ICE, we present exemplary results on the photoprocessing of undecane (C$_{11}$H$_{24}$) at 14 K as a feasibility study of the UV irradiation of aliphatic hydrocarbons. Aliphatics as part of the carbonaceous cosmic dust have been identified to be widespread in space by its well-known IR absorption bands at wavelengths of 3.4 $\mu$m, 6.8 $\mu$m and 7.3 $\mu$m \cite{chiar00,pendleton02,gunay20}. In addition, aliphatics are known to be present in cometary dust particles\cite{keller06}, and, recently, aliphatic hydrocarbons have been detected \textit{in-situ} by the Rosetta mission on comet 67P/Churyumov-Gerasimenko\cite{raponi20}, including short \textit{n}-alkanes (4-5 carbon atoms)\cite{schuhmann19}.

\section{The INFRA-ICE experimental station}

\begin{figure*}
    \centering
        \includegraphics[width=1\linewidth]{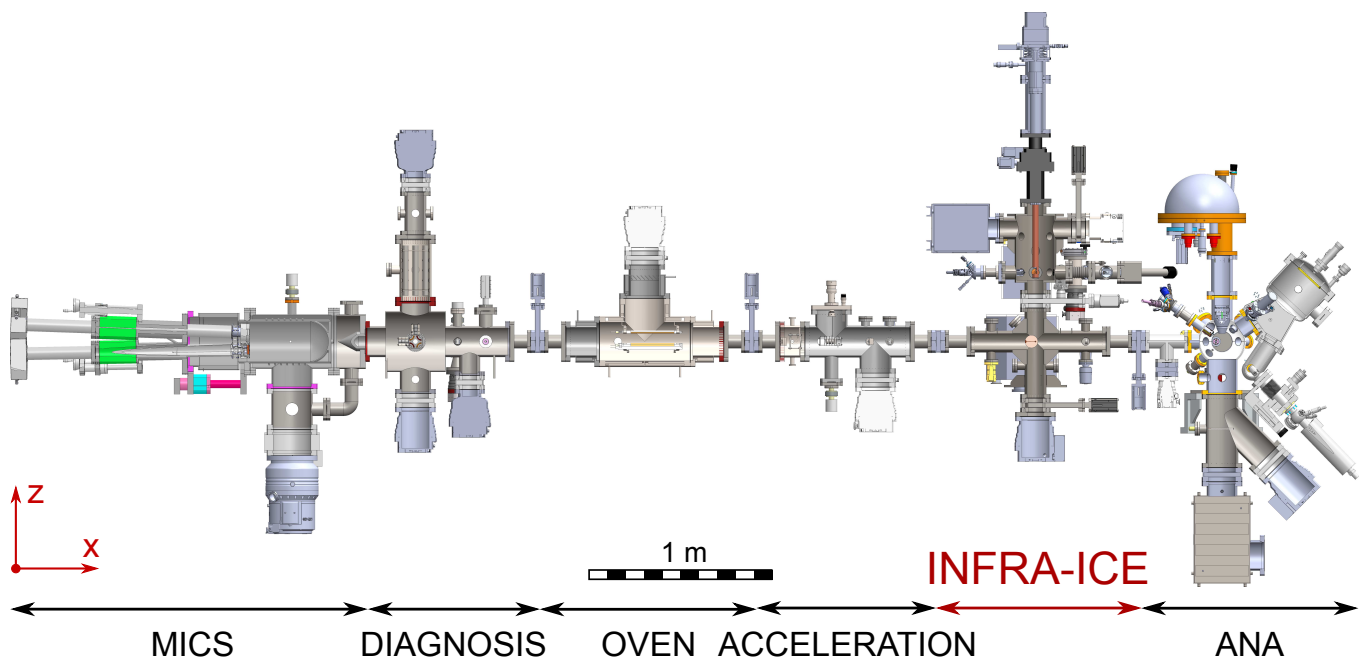}
        \caption{\label{fig:santorof1_rev1}Side sectional view of the \textit{Stardust} machine in its standard configuration. The beam of cosmic dust analogs travels along the positive x direction.}
\end{figure*}

The INFRA-ICE experimental station is a new ultra-high vacuum (UHV) module that has been incorporated into the \textit{Stardust} machine. A detailed description of the other \textit{Stardust} modules can be found elsewhere \cite{martinez18}. Briefly, the \textit{Stardust} machine (Fig. \ref{fig:santorof1_rev1}) is devoted to simulate in the laboratory the formation of cosmic dust in the atmosphere of AGB stars and comprises a set of different UHV modules that can be arranged in the most favorable configuration depending on the particular experimental requirements. The first module (MICS) consists of a multiple ion cluster source \cite{haberland91,martinez12} that allocates three independent magnetron sputter sources. The cosmic dust analogs are synthesized in this module and a beam of nanometer-size particles is produced, which travels along the machine. In the standard configuration of \textit{Stardust}, a module for beam diagnosis is located at the MICS exit. This module (DIAGNOSIS) is dedicated to the characterization of the particle beam properties and accommodates a Faraday cup, a quartz crystal microbalance and a quadrupole mass spectrometer with a mass range from 0 to 10$^6$ amu, which measure the charge, the production rate and the mass of the analogs, respectively. The next module (OVEN) is devoted to in-flight heating of the particle beam to temperatures of up to 1400 K via three 2 kW infrared lamps. Subsequently, \textit{Stardust} includes a module (ACCELERATION) to simulate the radiation pressure to which the dust analogs are subjected in the atmosphere of AGBs. This is achieved by ion optics whereby the nanometric dust analogs are ionized, accelerated and focused by an electron impact ionizer and a set of Einzel lenses. The INFRA-ICE module in the standard configuration of \textit{Stardust} is located between this module and an analysis module (ANA) in which the dust analogs are collected and can be analyzed by electron spectroscopies (X-ray Photoelectron Spectroscopy, Auger spectroscopy and Ultraviolet Photoelectron Spectroscopy) and Temperature Programed Desorption (TPD).

\begin{figure*}
    \centering
        \includegraphics[width=1\linewidth]{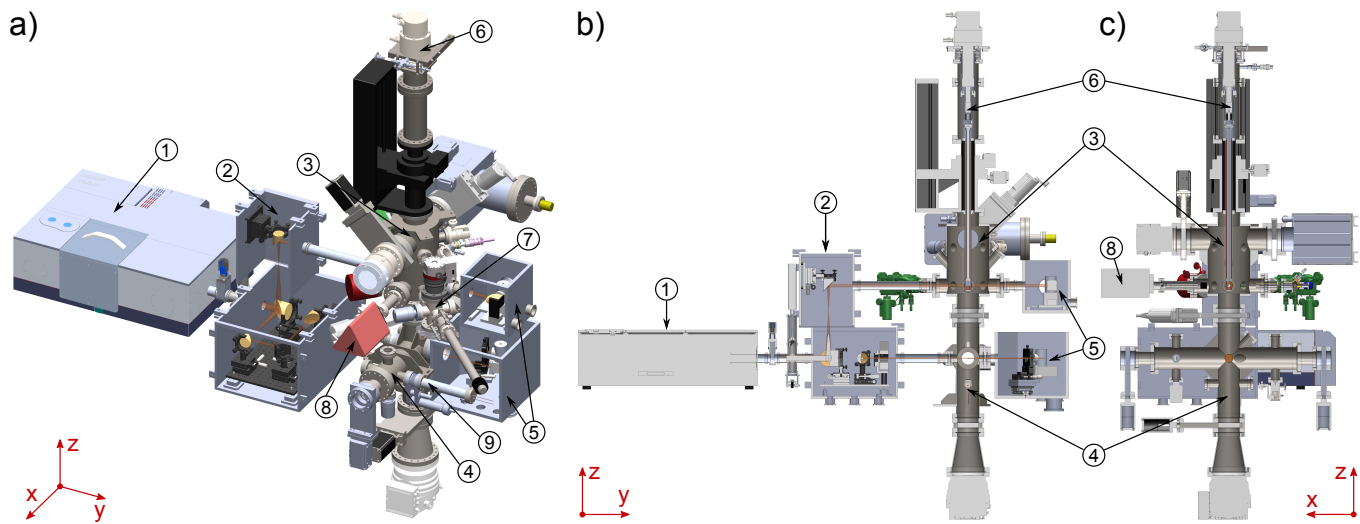}
        \caption{\label{fig:santorof2_rev1}a) Isometric view of the INFRA-ICE module. b) Front and c) side sectional views. The beam of cosmic dust analogs travels along the positive x direction. Key: 1.- IR spectrometer; 2.- IR coupling optics; 3.- Upper UHV chamber; 4.- Lower UHV chamber; 5.- Coupling optics for IR detectors; 6.- UHV close-cycle helium cryostat; 7.- Load-lock sample transfer chamber; 8.- Quadrupole mass spectrometer; 9.- Quartz crystal microbalance.}
\end{figure*}

The INFRA-ICE module (base pressure: 1 $\times$ 10$^{-10}$ mbar; at low temperature: 5 $\times$ 10$^{-11}$ mbar) is depicted in Figure \ref{fig:santorof2_rev1}a-c and a photograph of the setup is shown in Figure 3. It is a versatile setup that comprises two vertically connected UHV chambers. The lower one is on-axis with the beam of dust analogs produced in \textit{Stardust}, whereas the upper one lies inmediately above. Both chambers can be isolated by a UHV gate-valve and have independent vacuum equipment (Lower chamber: Bayard-Alpert gauge with tungsten filaments (Lewvac, UK) and turbomolecular pump \textit{HiPace 800} (Pfeiffer Vacuum GmbH, Germany); Upper chamber: Bayard-Alpert gauge with tungsten filaments (Lewvac, UK), turbomolecular pump \textit{HiPace 300} (Pfeiffer Vacuum GmbH, Germany) and ion pump \textit{VacIon Plus 150} (Agilent, USA) equipped with a titanium sublimation pump and a cryopanel). Thus, independent experiments can be carried out in the upper UHV chamber without interrupting the operation of the rest of the \textit{Stardust} machine. In addition, the lower chamber has UHV gate-valves at both the beam entrance and exit, allowing it to be isolated from the rest of the \textit{Stardust} machine without interrupting the UHV conditions, and operated independently. In this way, the INFRA-ICE module constitutes an autonomous experimental station.

On top of the upper chamber, a UHV close-cycle helium cryostat \textit{CCS-UHV/204} (Janis Research, USA) is mounted on a motorized 4-axis UHV manipulator (Huntington Mechanical Labs Inc., USA) comprising three linear (xyz) and one rotation (r) stages with resolutions of 10 $\mu$m and 0.1$\degree$, respectively. The r-axis has $\pm$ 180$\degree$ travelling range whereas the x- and y-axis have  $\pm$ 9 mm travelling range. The travelling range of the z-axis is 400 mm allowing for transferring  the sample from the upper to the lower UHV chambers and \textit{vice versa}. This enables to collect the cosmic dust analogs in the lower chamber (that is on axis with the dust analogs beam) and either analyze them by IR spectroscopy, which is the main characterization technique of the INFRA-ICE module, directly at this position, or transfer them to the upper chamber where energetic processing can be performed (see below).

\begin{figure}
    \centering
        \includegraphics[width=1\linewidth]{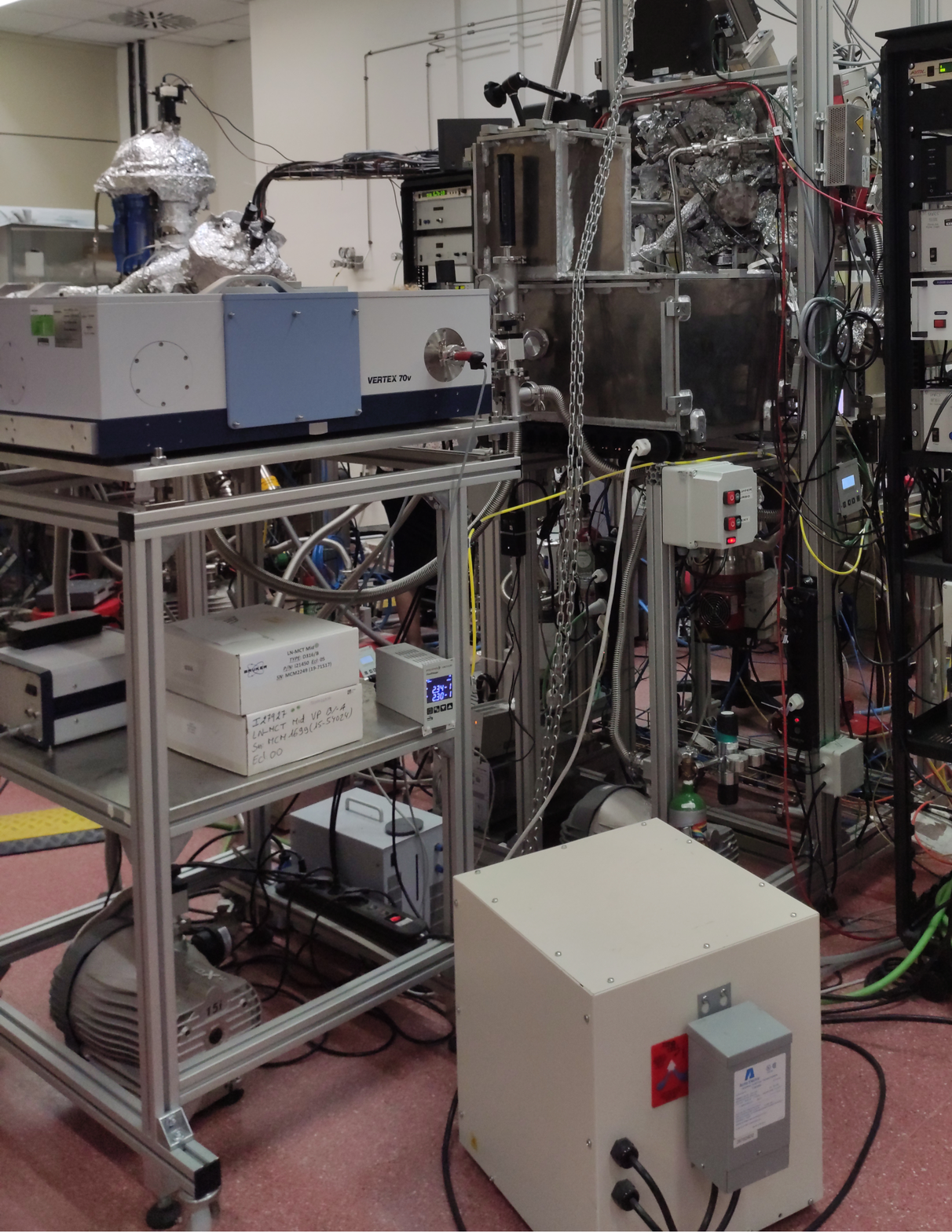}
        \caption{\label{fig:santorof3_rev1}Photograph of the INFRA-ICE module.}
\end{figure}

\begin{figure}
    \centering
        \includegraphics[width=1\linewidth]{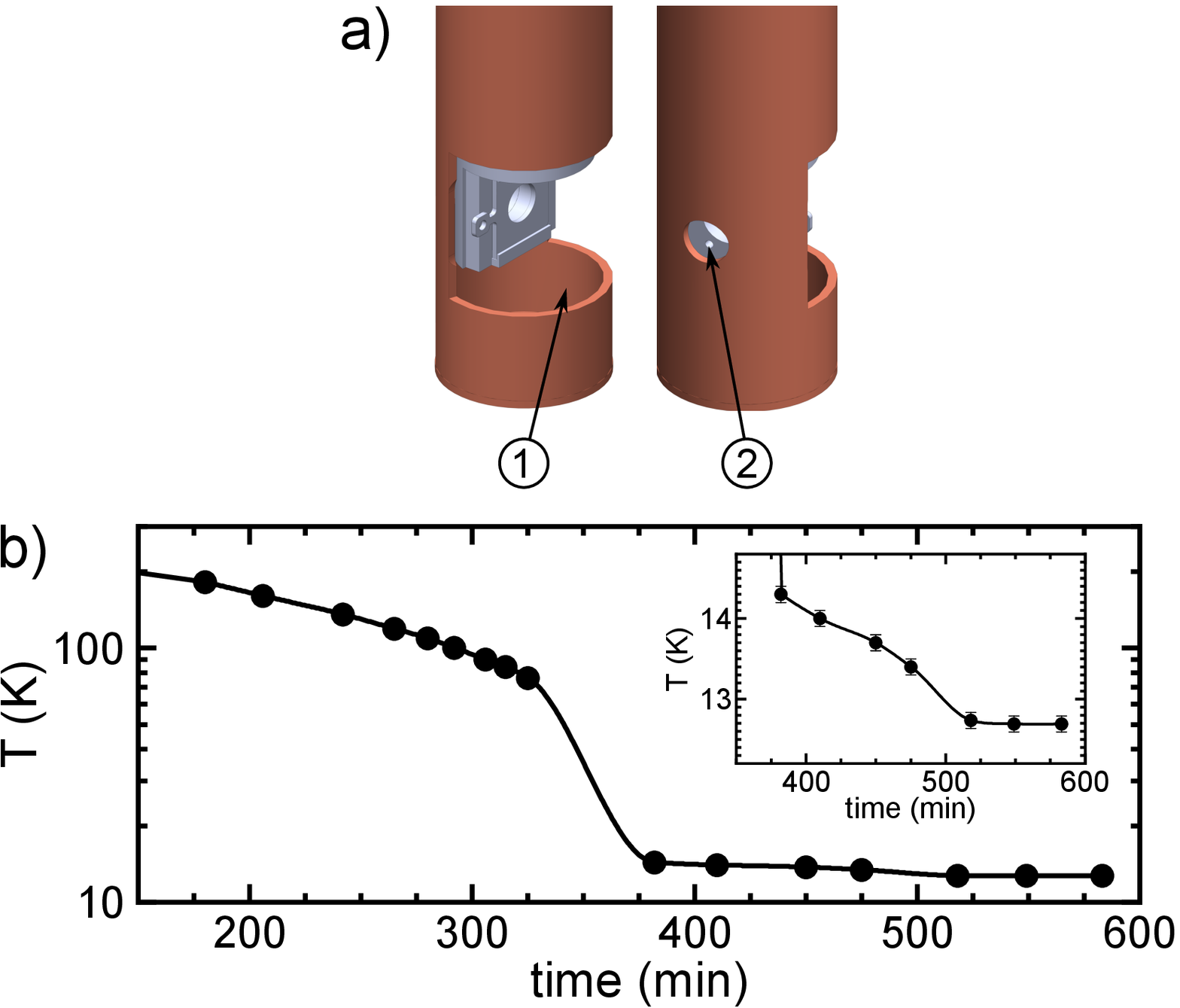}
        \caption{\label{fig:santorof4_rev1}a) Radiation shield for the cryostat cold finger at the sample position. Key: 1.- Front radiation shield optical access; 2.- Back radiation shield optical access. b) Temperature evolution at the sample position during cooling down at the maximum cooling power of the cryostat. The inset shows a zoom of the curve.}
\end{figure}

The cryostat operates in the temperature range 13-300 K at the sample position with a temperature stability of 0.1 K. Two Si-diode temperature sensors at different positions (at the sample position and 30 mm above) are used for temperature monitoring, and the temperature is controlled by a \textit{LakeShore 335} (Lake Shore Cryotronics Inc., USA) cryogenic temperature controller. In order to perform transmission and reflectance IR spectroscopy, a modified radiation shield and sample holder have been fabricated. (Fig. \ref{fig:santorof4_rev1}a). In particular, a sample holder with a 9 mm diameter hole is used and a cylindrical section has been drilled in the radiation shield permitting the IR illumination of the sample at the shallow angles needed for performing Infrared Reflection-Absorption Spectroscopy (IRRAS). In addition, a 9 mm diameter hole has been drilled in the back side of the radiation shield for IR spectroscopy in transmission. The maximum total power irradiated on the sample by the IR sources of the spectrometer has been estimated as 15 mW, which represents a very small heat load and does not affect the temperature stability. These modifications on the radiation shield and sample receiver increase the minimum temperature (T = 12.6 K) that can be achieved with respect to the cryostat specifications (temperature range according to specifications: 10-300 K). In addition, the long cold finger (length: 780 mm) needed to have enough travelling range from the upper to the lower chamber, notably increases the cooling time at the sample position (Fig. \ref{fig:santorof4_rev1}b).

\begin{figure}
    \centering
        \includegraphics[width=1\linewidth]{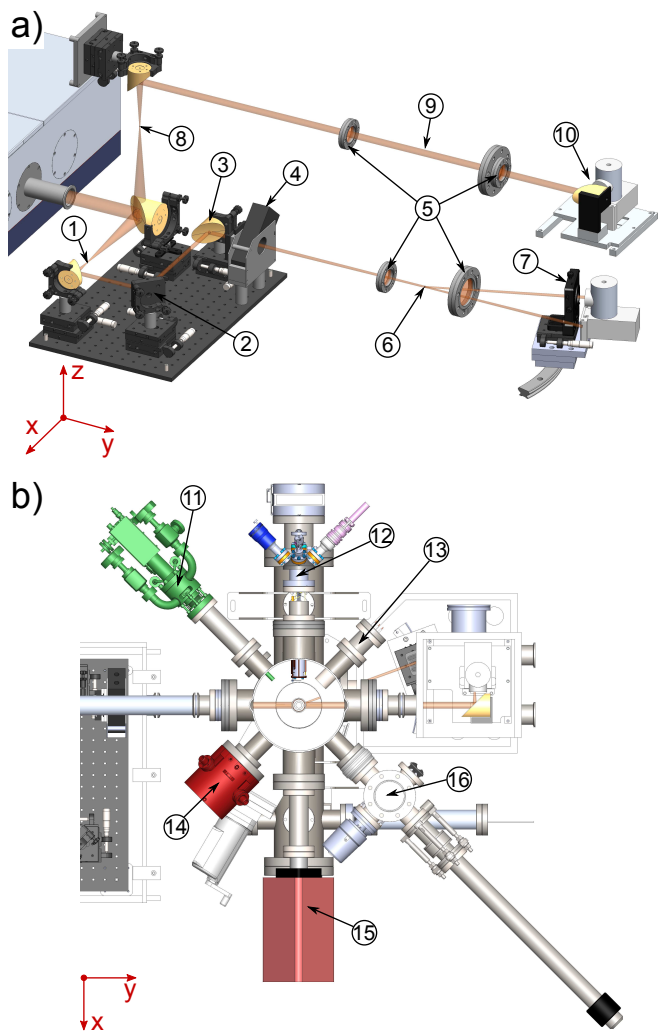}
        \caption{\label{fig:santorof5_rev1}a) Radiation shield for the cryostat cold finger at the sample position. b) Top sectional view of the upper chamber. Key: 1.- 4.5$\times$ beam compressor for the lower UHV chamber; 2.- Flat mirror; 3.- Focusing mirror for the lower UHV chamber; 4.- Polarizer and photoelastic modulator; 5.- Wedged ZnSe windows; 6.- Sample position in the lower UHV chamber. To represent the optical paths for transmission and reflectance spectroscopy, the IR beam has been split at this position; 7.- Focusing ZnSe lens and IR detector; 8.- 2.25$\times$ Beam compressor for the upper UHV chamber; 9.- Sample position in the upper UHV chamber. 10.- Focusing mirror and IR detector; 11.- UV source; 12.- Hydrogen cracker; 13.- Electron gun; 14.- Ion source; 15.- Quadrupole mass spectrometer; 16.- Load-lock sample transfer chamber.}
\end{figure}

IR spectroscopy can be performed in the lower chamber in transmission and reflectance modes (IRRAS), or in the upper chamber in transmission. This is achieved by a properly designed optical path to guide the IR beam from one of the exit ports of the spectrometer to the sample positions inside the chambers (Fig. \ref{fig:santorof5_rev1}a). All the coupling optical elements are held under vacuum (10$^{-1}$ mbar) to suppress contributions from atmospheric water and carbon dioxide. To isolate the spectrometer and coupling optics from the UHV of the chambers, we used ZnSe windows (wedged: 5.8 mrad) bonded to double-sided CF flanges.

The first element in the optical setup comprises a \textit{Vertex 70V} Fourier Transform IR (FT-IR) spectrometer (Bruker Optik GmbH, Germany) with both mid-infrared (MIR) and near infrared (NIR) capability, covering a spectral range from 12800 cm$^{-1}$ to 400 cm$^{-1}$ (0.78 to 25 $\mu$m) with a maximum spectral resolution of 0.16 cm$^{-1}$. However, the ZnSe windows present a transmission cut-off starting at 650 cm$^{-1}$ with no transmission below 500 cm$^{-1}$, which sets the upper wavelength limit of the spectral range. The Mid-Infrared (MIR) configuration of the spectrometer comprises two sources (a standard air-cooled globar MIR source and a water-cooled high-power MIR source), a KBr beamsplitter and two detectors (a DLaTGS detector and a liquid nitrogen cooled Mercury-Cadmium-Telluride (MCT) detector). On the other hand, the NIR configuration consists of a tungsten halogen lamp, a CaF$_2$ beamsplitter and an InGaAs-diode detector. A rapid-scan option for performing time-resolved spectroscopy is available at a maximum scanner velocity of 160 kHz.

The optical path (Fig. \ref{fig:santorof5_rev1}a) towards the lower UHV chamber consists in, firstly, a 4.5$\times$ beam compressor that comprises two 90$\degree$ off-axis parabolic gold mirrors with focal lengths of 9” and 2”, respectively. Due to the shallow angles used for illumination in IRRAS, overillumination (i.e., beam size at sample position larger than sample size) is typically found. Thus, this beam compressor provides a better matching between the IR spot size at the sample position and the sample size (standard sample size 10 $\times$ 10 mm$^2$). The compressed beam is then reflected by a flat gold mirror that directs the IR beam to a 90$\degree$ off-axis gold parabolic mirror with a focal length of 500 mm, which focuses the IR beam at the sample position in the center of the defined dust particle beam axis in the lower UHV chamber. In between this mirror and the ZnSe UHV entrance window, an IR polarizer and a ZnSe photoelastic modulator (PEM) (Hinds Instruments, OR, USA) operating at 50 kHz can be installed enabling Polarization-Modulated IRRAS (PM-IRRAS) measurements to be performed. A detailed description of the PM-IRRAS technique can be found elsewhere \cite{frey06}. Once the sample is illuminated, the IR beam, either transmitted o reflected, goes through a ZnSe UHV exit window that limits the minimum incident angle for IRRAS to 76$\degree$ with respect to the sample surface normal (note that the incident beam is fixed whereas the incident angle is selected by rotating the sample). Then a 40 mm diameter ZnSe lens is used to focus the IR beam onto the detector. A lens is used instead of a parabolic mirror to maintain the polarization properties of the IR beam required for performing PM-IRRAS. The IR detector and the lens are mounted on a curved guideway to accurately position the detector with respect to the reflected IR beam whilst maintaining the same sample-lens distance.

To access the upper UHV chamber, the first off-axis parabolic mirror after the spectrometer is rotated by 90 degrees to guide the IR beam upwards and a 2.25$\times$ beam compressor is found employing a second 90$\degree$ off-axis parabolic gold mirror of 4” focal length. This beam compressor preserves the beam collimation at the exit port of the spectrometer while reducing the spot size to better adjust to the sample size and to the modifications performed on the cryostat sample receiver and radiation shield (see below). The compressed IR beam crosses the ZnSe entrance window, is transmitted through the sample and exits the upper UHV chamber through another ZnSe window. Since in this chamber only transmission spectroscopy is performed, a 90$\degree$ off-axis parabolic gold mirror is used to focus the beam onto the detector.

To monitor the deposition rate of the dust analogs produced with \textit{Stardust}, the lower chamber is equipped with a quartz crystal microbalance (QCM). It is mounted on a UHV linear translator with the linear motion in the direction perpendicular to the NP beam. In this way, it can be placed in the NP beam axis to monitor the deposition rate and retracted to allow the NP beam to travel towards the next module of \textit{Stardust}.

The upper chamber allocates a quadrupole mass spectrometer (QMS) \textit{PrismaPlus QMG 220 M2} (Pfeiffer Vacuum GmbH, Germany) with a mass range of 1-200 amu. The QMS serves to check the residual gas in the UHV chamber prior to the deposition of molecular ices as well asto monitor the ice deposition process. It also allows temperature programmed desorption measurements as it has direct view of the sample surface.

In addition, the upper chamber also allocates several processing equipment (Fig. \ref{fig:santorof5_rev1}f), permitting the exposure of the samples to similar processing as those occurring in the interstellar clouds of the ISM. As mentioned in the previous section, the main energetic processes in molecular clouds are UV radiation and cosmic rays, the latter producing both ions and a cascade of secondary electrons.  The UV field in these regions is dominated by the Lyman-$\alpha$ emission of atomic hydrogen whereas the ions have energies in the keV-MeV range. Whilst high energy ions mainly promote the sputtering of the material\cite{johnson98}, ions in the keV range more efficiently lead to physico-chemical changes in the icy mantles that can result in the formation of new chemical species\cite{strazzulla01}. On the other hand, most of the large number of secondary electrons produced by cosmic ray impact on icy mantles and dust grains are low energy secondary electrons ($<$ 100 eV), which can induce a wide variety of radiation-driven chemical reactions\cite{boyer16,esmaili18,shulenberger19}. The processing equipment incorporated in INFRA-ICE is mainly dedicated to simulate the main processes inducing chemical changes in the ice-dust system in dense molecular clouds and therefore include: (i) a UV source \textit{40A2} (Prevac, Poland) that can be operated with different discharge gases (such as H$_2$, He or Ar), so that the UV source can provide different spectral emission profiles and photon energy depending on the discharge gas employed, (ii) an ion source  \textit{IQP 10/63} (Specs GmbH, Germany) providing ion energies from 0.2 to 6 keV, and (iii) an electron flood gun \textit{FG 15/40} (Specs GmbH, Germany) delivering electrons of energies between 1 and 500 eV. Moreover, in the diffuse clouds of the ISM, atomic hydrogen is particularly abundant and to simulate the exposure of cosmic dust to atomic hydrogen in these interstellar regions, the upper chamber also allocates a hydrogen thermal gas cracker \textit{TGC-H} (Specs GmbH, Germany) for the exposure of the analogs to atomic hydrogen.

Finally, for ice growth, the gases (either pure or gas mixtures) are supplied from a gas mixing system and injected through leak valves located both in the upper and lower chamber. The gas mixing system is thoroughly described in the Supplementary Material elsewhere\cite{martinez18}. Briefly, it can accommodate up to four gas bottles and two liquid reservoirs. Both the gases and liquid vapors are introduced in the mixing chamber (base pressure $<$ 10$^{-7}$ mbar) through calibrated gas-flow valves and a homogeneous gas mixture is achieved by in-flow mixing via a rotary pump equipped with a zeolite filter.

\section{Selected results: photochemistry of C$_{11}$H$_{24}$ at low temperature}

To illustrate some of the capabilities of the INFRA-ICE experimental station, in this section we present exemplary results on the photochemistry of undecane (C$_{11}$H$_{24}$)  at 14 K. Firstly, the calibration of the photon flux of the UV lamp employed is presented, which is needed to derive quantitative results for the photochemical experiments. 

\subsection{UV lamp flux calibration}

UV irradiation in dense molecular clouds produces a rich photochemistry of the icy mantles that are accreted on cosmic dust particles. The emission from H$_2$ discharge lamps has been shown to satisfactorily reproduce the radiation field of the diffuse interstellar medium \cite{jenniskens93} and are, therefore, commonly employed to simulate the UV photoprocessing of interstellar ice analogs. The vacuum UV emission of these lamps is dominated by the Lyman-alpha emission from atomic hydrogen ($\lambda$ = 121.6 nm) with contributions from molecular hydrogen emission centered at around $\lambda$ = 160 nm \cite{chen14}.

A common procedure for deriving the photon flux of hydrogen discharge lamps is O$_2$ actinometry \cite{baratta02, munozcaro10}. This measures the photochemical conversion of O$_2$ to O$_3$ and, through the quantum yield of the reaction, it is possible to convert the number of O$_3$ molecules produced per time unit into photon flux. Usually, the column density (number of molecules per unit area along the observation direction) of O$_3$ is derived from the ${\nu_3}$ asymmetric stretch of O$_3$ at 1040 cm$^{-1}$, whose band strength is known for the gas-phase.

However, this method presents several disadvantages. Firstly, O$_2$ has no IR absorption features and therefore only the increase in the ${\nu_3}$ mode of O$_3$ is observed during actinometric measurements. More importantly, even if solid-phase quantum yield and solid-phase band strength data for the ${\nu_3}$ mode of O$_3$ are reported in the literature \cite{fulvio14, teolis07}, these depend on the structure of the O$_2$ and O$_3$ ices. Finally, the optimum UV range in which O$_2$ photolysis takes place is from 130 nm to 190 nm \cite{chemicalactinometry}, which covers most of the UV emission of molecular hydrogen but excludes the Lyman-$\alpha$ line of atomic hydrogen.

For these reasons, other actinometric systems (e.g., the CO photoproduction from CO$_2$ \cite{martindomenech15}) have been employed, although for these systems solid-phase quantum-yield data are usually not available. The use of silicate photodiodes \cite{cottin03} and metallic meshes \cite{chen14} avoid these problems, but an accurate calibration of the devices in the vacuum UV region is needed, which requires an already calibrated lamp, not always available. A straightforward method for flux calibration was developed by Fulvio \textit{et al.} using the photocurrent generated in a gold photodetector \cite{fulvio14}.

To calibrate our UV lamp, we have used the photolysis of methane, which has been thoroughly investigated in ices and accurate photodestruction cross-section data in the far UV spectral range are available in the literature \cite{gerakines96, baratta02, cottin03}. This method relies on measuring the photolysis rate and is therefore independent of the band strength of the IR bands used; thus, it can even be applied to IR bands with unknown band strength.

The photolysis of an optically thin molecular ice can be described by a first-order reaction kinetics\cite{cottin03}
\begin{equation}
 			\frac{dn}{dt}=-kn
            \end{equation}
where $n$ is the number density of molecules and $k$ the photolysis rate, which is dependent on both the photodestruction cross-section $\sigma$($\lambda$) and the photon flux $I$($\lambda$). Assuming these two values to be constant over the wavelength range of interest, which is a reasonable approximation for narrow spectral ranges,\cite{cottin03} $k$ can be simply expressed as the product of $\sigma$ and $I$. Thus, by measuring the temporal evolution of the IR bands corresponding to the molecule being photolyzed, the incident photon flux is easily derived.

\begin{figure}
    \centering
        \includegraphics[width=1\linewidth]{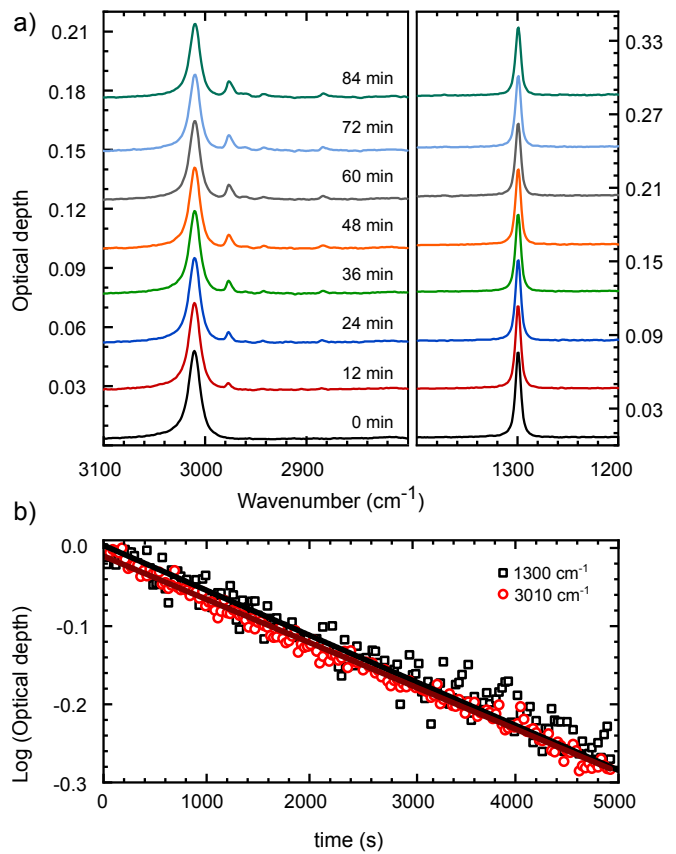}
        \caption{\label{fig:santorof6_rev1}a) IR absorption spectra during the photolysis of CH$_4$ ice. The irradiation time of each spectrum is indicated in the figure and curves have been shifted for clarity. b) Temporal evolution of the optical depth for the 1300 cm$^{-1}$ and 3010 cm$^{-1}$ bands. The solid lines correspond to the linear fit of the data.}
\end{figure}

Figure \ref{fig:santorof6_rev1}a shows the IR spectra of a CH$_4$ ice at different UV exposure times. The experiments were performed in the upper chamber of the INFRA-ICE experimental station. The CH$_4$ ice was deposited at 14 K on IR transparent KBr substrates through the so-called background deposition method \cite{accolla11}. In our discharge lamp, the UV emission is guided through a windowless quartz capillary. Windowless discharge lamps have the advantage of avoiding photon flux losses due to the deposition on the windows of material generated in the discharge , a phenomenon that is commonly observed when working with windowed discharge lamps\cite{jenniskens93}. These deposits produce an increase in the absorption of the window over time and, therefore, a decrease of the photon flux on the sample during irradiation\cite{cottin03}. However, in the case of windowless lamps a flow of the discharge gas is introduced into the chamber, which might interact with the sample. For the photolysis experiments, our UV lamp was operated with a flow of H$_2$ (purity 99.99 \%) that resulted in a pressure in the upper chamber of INFRA-ICE of 2 $\times$ 10$^{-8}$ mbar. Therefore, the experiments were performed in an enriched H$_2$ atmosphere but at 14 K H$_2$ does not condense and, thus, H$_2$ was not deposited on the sample. According to the manufacturer, these H$_2$ flow conditions maximize the emission of the Lyman-$\alpha$ line over the emission of molecular hydrogen. A power of 60 W was applied to the discharge. During the UV exposure, IR spectra were concurrently recorded every 30 s with a spectral resolution of 2 cm$^{-1}$ and 64 scans were coadded for each spectrum.

The spectra of pure CH$_4$ ice present two distinct absorption bands at 1300 cm$^{-1}$ and 3010 cm$^{-1}$, which correspond to the C-H bending and C-H stretching vibrational modes of CH$_4$, respectively. As the ice is irradiated, both bands decrease in intensity and new bands appear in the range 3000-2800 cm$^{-1}$, which are due to the CH$_2$ and CH$_3$ stretching modes of C$_2$H$_6$ and C$_3$H$_8$. The production of larger alkanes from the vacuum UV exposure of CH$_4$ is a well-known process involving the photolysis of CH$_4$ to CH$_2$ + H$_2$ and to CH$_3$ + H \cite{gerakines96}.

Figure \ref{fig:santorof6_rev1}b, shows the temporal evolution of the natural logarithm of the integrated optical depth (normalized to the maximum value) of the 1300 cm$^{-1}$ and 3010 cm$^{-1}$ bands of CH$_4$ during UV exposure. Linear fittings to the data yield photolysis rates of (5.76 $\pm$ 0.07) $\times$ 10$^{-5}$ s$^{-1}$ and (5.53 $\pm$0.04) $\times$10$^{-5}$ s$^{-1}$ for the 1300 cm$^{-1}$ and 3010 cm$^{-1}$ bands, respectively. These correspond to a photon flux (integrated over the complete UV spectral range) of (6.2 $\pm$ 0.6) $\times$ 10$^{14}$ ph s$^{-1}$ cm$^{-2}$, using a photodestruction cross section of 9.1 $\times$ 10$^{-20}$ cm$^2$ for CH$_4$ \cite{cottin03} and considering the mean value of both derived photolysis rates.

\subsection{UV photochemistry of C$_{11}$H$_{24}$}

Recently, by simulating the circumstellar envelope of carbon-rich AGBs in the laboratory with the \textit{Stardust} machine, we have shown that the interaction of atomic carbon with hydrogen, the latter being the most abundant gaseous species in AGBs, leads predominantly to aliphatics, including alkanes \cite{martinez20}. Moreover, as mentioned in the introduction, aliphatics are widespread in space as component of the carbonaceous cosmic dust\cite{pendleton02,gunay20} and are also present in cometary dust particles\cite{keller06}. Also recently, aliphatic hydrocarbons have been identified \textit{in-situ} by the Rosetta mission in the comet 67P/Churyumov-Gerasimenko\cite{raponi20}, including short \textit{n}-alkanes of 4-5 carbon atoms in the gas-phase\cite{schuhmann19}. In this section, to show some of the capabilities of INFRA-ICE, we present the results obtained during the UV exposure of undecane (C$_{11}$H$_{24}$) as a feasibility study of the UV photochemistry of aliphatic hydrocarbons at low temperatures.

The experiments were performed in the upper chamber of INFRA-ICE, where C$_{11}$H$_{24}$ vapor was deposited at 14 K on KBr substrates and exposed to UV radiation for 240 min. For the deposition C$_{11}$H$_{24}$ (purity: $\geqslant$ 99 $\%$)  was loaded into a Pyrex ampoule with conflat fittings and warmed to 345 K. The UV lamp was loaded with H$_2$ (purity: 99.99 $\%$) and the hydrogen discharge was carried out under the same conditions as those described in the previous section. Thus, the total UV exposure corresponded to a photon fluence of ca. 9 $\times$ 10$^{18}$ photons cm$^{-2}$. Throughout the complete UV treatment, IR spectra were simultaneously acquired every 190 s in transmission mode with a spectral resolution of 2 cm$^{-1}$, coadding 128 scans for each spectrum.

\begin{figure}
    \centering
        \includegraphics[width=1\linewidth]{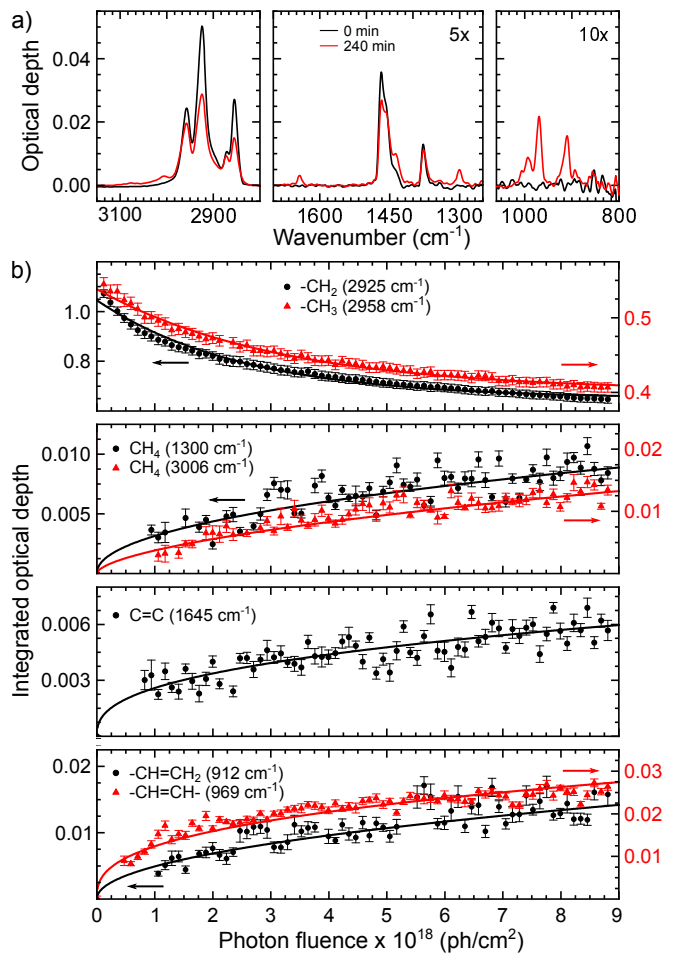}
        \caption{\label{fig:santorof7_rev1}a) IR spectra of C$_{11}$H$_{24}$ as deposited and after 240 min of UV exposure. b) Evolution of the optical depth of selected IR bands during UV exposure. The arrows indicate the y-axis for each curve whereas the solid lines are guides to the eye.}
\end{figure}

The IR spectra of C$_{11}$H$_{24}$, and of \textit{n}-alkanes and aliphatics in general, are dominated by the absorptions in the spectral region 3000-2800 cm$^{-1}$, which correspond to the CH$_2$ and CH$_3$ stretching modes (see Fig.\ref{fig:santorof7_rev1}a) . Additionally, the bands at around 1380 cm$^{-1}$ along with those at around 1460 cm$^{-1}$ are ascribed to the highly distinctive CH$_3$ symmetric bending ("umbrella" deformation) mode and to the CH$_3$ asymmetric bending/CH$_2$ scissoring modes, respectively, and are very characteristic of aliphatics. In the case of \textit{n}-alkanes, the position of the CH$_3$ symmetric bending mode and the band structure of the absorptions at around 1460 cm$^{-1}$ are highly sensitive to the conformational structure of the molecules \cite{snyder78} and are therefore very sensitive to crystallinity. For \textit{n}-alkanes, the crystalline phase is comprised exclusively of CH$_2$-CH$_2$ \textit{trans} conformers, whereas \textit{gauche} conformers are associated with amorphous material.

In our case, the IR spectrum of as-deposited C$_{11}$H$_{24}$ reflects the amorphous structure of the material as can be derived from the shape, position and width of the bands at 1378 cm$^{-1}$ and in the 1420-1480 cm$^{-1}$ region \cite{snyder63,snyder78}. The amorphous structure of C$_{11}$H$_{24}$ is related to the temperature of the KBr substrate were the ice is grown. Figure \ref{fig:santorof7_rev1}a shows the IR spectra of non-irradiated C$_{11}$H$_{24}$ and after an UV exposure of 240 min. Clear changes are observed after the UV irradiation, which consist mainly in a reduction in absorption in the CH$_2$ and CH$_3$ stretching modes and the appearance of new absorption bands. The positions and assignments of the observed IR bands are listed in Table \ref{tab:table1}. The evolution of some selected IR bands during the UV photoprocessing of C$_{11}$H$_{24}$ is shown in Figure \ref{fig:santorof7_rev1}b. The integrated optical depth for each absorption feature has been obtained by band deconvolution using Gaussian curves.

\begin{table}
\caption{\label{tab:table1}IR band assignment}
\centering

\begin{tabular}{ccl}
\hline\hline 
Wavenumber &Wavelength &Assignment\footnote{The vibrational modes are abbreviated as: $\nu$	: stretching; $\delta$: deformation (b: bend, sc: scissor); $\gamma$: wagging; s: symmetric; as: assymetric; oop: out-of-plane.} \footnote{Assignments from \cite{snyder63,snyder78,gerakines96,socrates}}\\
cm$^{-1}$ & $\mu$m & \\
\hline

\multicolumn{3}{c}{As-deposited C$_{11}$H$_{24}$}  \\
2957	&	3.38	&	$\nu_{as}$ CH (CH$_3$)\\
2925	&	3.42	&	$\nu_{as}$ CH (CH$_2$)\\
2871	&	3.48	&	$\nu_{s}$ CH (CH$_3$)\\
2854	&	3.50	&	$\nu_{s}$ CH (CH$_3$)\\
1470	&	6.80	&	$\delta_b$ CH (CH$_2$)\\
1458	&	6.86	&	$\delta_{sc}$ CH (CH$_2$), $\delta_{as}$ CH (CH$_3$)\\
1437	&	6.96	&	$\delta_{sc}$ CH (CH$_2$), $\delta_{as}$ CH (CH$_3$)\\
1378	&	7.26	&	$\delta_{s}$ CH (CH$_3$)\\
\hline

\multicolumn{3}{c}{New bands after UV processing of C$_{11}$H$_{24}$} \\
3076	&	3.25	&	$\nu_{as}$ CH (=C-H)\\
3006	&	3.33	&	$\nu_{as}$ CH (CH$_4$)\\
1645	&	6.08	&	$\nu$ C=C\\
1300	&	7.69	&	$\delta$ CH (CH$_4$)\\
994		&	10.06	&	$\gamma_{oop}$ CH (-CH=CH$_2$) $_{vinyl}$\\
967		&	10.34	&	$\gamma_{oop}$ CH (-HC=CH-) $_{trans}$\\
912		&	10.96	&	$\gamma_{oop}$ CH (-CH=CH$_2$) $_{vinyl}$\\
\hline\hline 

\end{tabular}
\end{table}

It can be observed that the decrease in intensity of the CH$_2$ and CH$_3$ stretching modes follows an exponential decay corresponding to first order reaction kinetics as expected for a photolysis process (the symmetric CH$_2$ and CH$_3$ stretching modes at 2925 cm$^{-1}$ and 2958 cm$^{-1}$, respectively, are shown in Fig. 5b). Additionally, the evolution of the bands at around 1460 cm$^{-1}$ presents a complex behavior. In particular, the band at 1470 cm$^{-1}$, which can be ascribed to CH$_2$ moieties, decreases in the same way as the CH$_3$ and CH$_2$ stretching modes. On the other hand, the shoulder at 1437 cm$^{-1}$ is observed to increase. This might be related to the formation of unsaturated hydrocarbon moieties (see below) since a band at around 1440 cm$^{-1}$ can be assigned to the bending mode of methylene moieties in the presence of adjacent unsaturated groups \cite{socrates}. However, a deep interpretation of the spectral changes observed in this spectral region is difficult since, in addition to the formation of new chemical species, the UV exposure of alkanes can induce a conformational rearrangement that produces changes in the band structure of the absorptions at around 1460 cm$^{-1}$, which, as abovementioned, is very sensitive to the conformational structure.

On the other hand, the new IR bands at 1300 cm$^{-1}$ and 3006 cm$^{-1}$ reveal the production of methane. The release of CH$_2$ and CH$_3$ from the photolysis of C$_{11}$H$_{24}$ promotes the interaction of these with H$_2$ and H (both released from the photolysis of undecane and from the H$_2$ in the chamber) to form CH$_4$, a process that is well-known for the photolysis of alkanes \cite{vuv_photochemistry}. More interestingly, as the UV irradiation proceeds, a new infrared band appears at 1645 cm$^{-1}$, which is assigned to the C=C stretching mode of alkenes \cite{socrates}, proving the formation of unsaturated hydrocarbons. In addition, a very weak band at 3076 cm$^{-1}$ can be observed in the spectrum of the UV irradiated material in Figure \ref{fig:santorof7_rev1}a. This band is assigned to the CH stretching of olefins \cite{socrates}.

The nature of the alkenes formed by photoprocessing is revealed by the absorption features in the spectral region corresponding to deformation vibrations of C-H (1000-800 cm$^{-1}$). In particular, the bands at 994 cm$^{-1}$ and 912 cm$^{-1}$ are \sout{very} characteristic of vinyl moieties whereas the band at 967 cm$^{-1}$ is characteristic of \textit{trans} vinylene hydrocarbons \cite{socrates}. Therefore, the UV photoprocessing of C$_{11}$H$_{24}$ promotes the formation of olefinic moieties both at the end and in the back-bone of the chains in the form of vinyl (-CH=CH$_2$) and vinylene chemical groups (-CH=CH-), respectively.

The formation of CH$_4$ and unsaturated hydrocarbons is further confirmed by TPD measurements during the sublimation of the UV-exposed C$_{11}$H$_{24}$ ice. To conduct the TPD measurements two different procedures were employed. An initial TPD measurement was performed using multiple ion detection at selected m/z signals whereas a second TPD measurement was carried out acquiring the complete mass spectra of the desorbed gases in the range m/z = 0-200. The latter allowed for a deeper characterization of the desorbed chemical species. Both procedures were performed on identical UV-exposed C$_{11}$H$_{24}$ samples (UV fluence of ca. 9 $\times$ 10$^{18}$ ph cm$^{-2}$) and, for comparison purposes, on identical C$_{11}$H$_{24}$ deposits without UV exposure. A heating rate of 5 K min$^{-1}$ was used in all cases.

\begin{figure}
    \centering
        \includegraphics[width=1\linewidth]{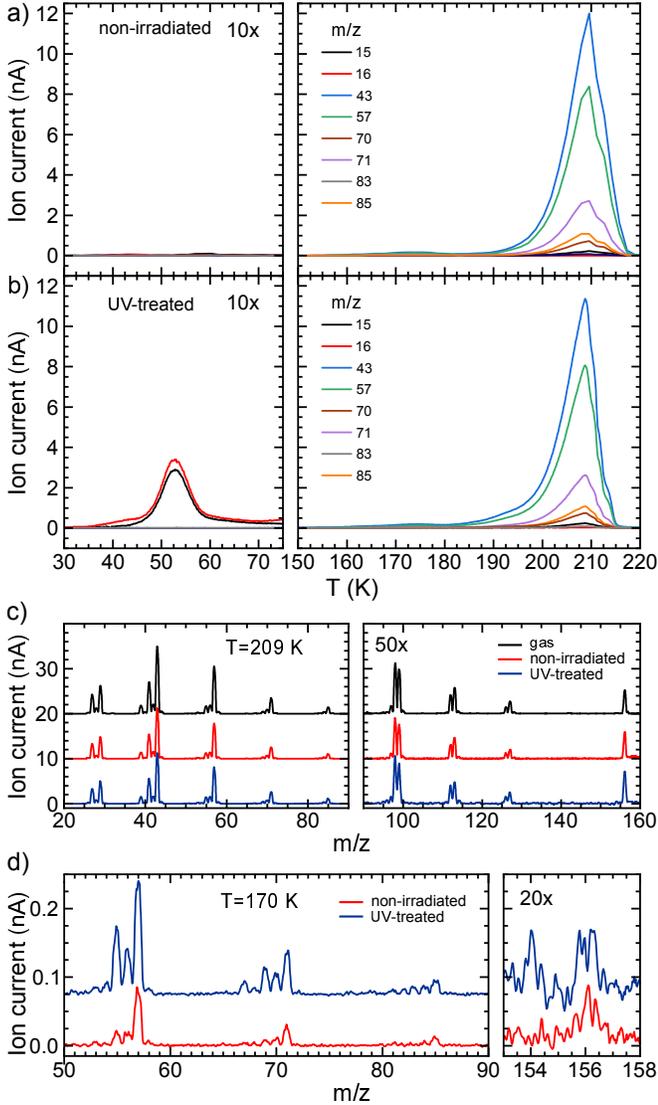}
        \caption{\label{fig:santorof8_rev1}a) Evolution of some selected m/z peaks during the TPD of non-irradiated C$_{11}$H$_{24}$ and b) after 240 min of UV exposure. c) Mass spectra of the species desorbed at 209 K for the as-deposited and UV exposed C$_{11}$H$_{24}$. For comparison, the mass spectrum of C$_{11}$H$_{24}$ in gas phase is included. The curves have been vertically shifted for clarity.  d) Mass spectra of the species desorbed at 170 K for the non-irradiated and UV treated C$_{11}$H$_{24}$. The curves have been vertically shifted for clarity.}
\end{figure}

Figure \ref{fig:santorof8_rev1}a and b show the evolution of the signal of some selected m/z values during the thermal desorption of neat and UV-exposed C$_{11}$H$_{24}$ molecular ices, respectively. The m/z values were selected according to the electron-impact dissociation patterns of the chemical species that were expected to be formed after the UV processing. In particular, m/z = 15 and 16 are characteristic of CH$_4$, m/z = 43, 57, 71 and 85 correspond to the strongest signals of C$_{11}$H$_{24}$ and m/z = 70 and 83 are characteristic for long chain alkenes \cite{nist}, though they also present contributions from C$_{11}$H$_{24}$. Except the clear desorption of CH$_4$ with a maximum at around 53 K, no evidences of new chemical species were detected by the multiple ion detection procedure, not even by comparison with a non-irradiatedC$_{11}$H$_{24}$ ice. In fact, the mass spectra of the unprocessed and UV-exposed samples at 209 K (which corresponds to the temperature of the maximum desorption of C$_{11}$H$_{24}$) are identical and resemble that of C$_{11}$H$_{24}$ in gas-phase (Fig. \ref{fig:santorof8_rev1}c).

However, important changes can be observed at 170 K (Fig.\ref{fig:santorof8_rev1}d). This is the temperature at which the desorption of C$_{11}$H$_{24}$ begins and, therefore, at this temperature the signal coming from new chemical compounds (that desorb in the same temperature range and with mass peaks overlapping with those of undecane) is not obscured by the much more intense signal from the desorption of C$_{11}$H$_{24}$ at higher temperatures. In comparison to the non-irradiated C$_{11}$H$_{24}$ ice, the mass spectrum of the UV-treated C$_{11}$H$_{24}$ exhibit clear differences at 170K in the regions at m/z = 53-57, m/z = 67-71 and m/z = 81-85, among others. In particular, the peak structure of the mass spectrum of the UV-treated ice, present in these regions fragmentation patterns typical of alkenes \cite{nist} for which the maxima in the electron impact fragmentation pattern is lowered in steps by m/z = 1 and 2 regarding the fully saturated counterpart\cite{nist}. Finally, a very faint peak can be appreciated at m/z = 154 which can be ascribed to C$_{11}$H$_{22}$\cite{nist}.

\section{Conclusions}

We have presented a new experimental station devoted to simulate in the laboratory the complex conditions of the coldest regions of the ISM and to investigate the interaction of cosmic dust with ices of astrophysical interest. The INFRA-ICE experimental station is a versatile UHV setup that can be operated independently or integrated as a module of the \textit{Stardust} machine, which is devoted to simulate in the laboratory the formation and evolution of cosmic dust in the circumstellar envelopes of asymptotic giant branch stars. When incorporated into \textit{Stardust}, INFRA-ICE expands the capabilities of \textit{Stardust} allowing for the simulation in the laboratory of the complete journey of cosmic dust from its formation in evolved stars to its processing in the densest regions of the ISM, where cosmic dust is coated with molecular ices. The main analysis technique of  INFRA-ICE is IR spectroscopy (both transmission and reflectance) and a quadrupole mass spectrometer with direct view of the sample can be employed to perform Temperature Programmed Desorption (TPD) measurements. In addition, a set of processing equipment (UV source, ion gun, electron gun and hydrogen gas cracker) is available for exposing the samples to similar processes as those encountered in dense molecular clouds.

As an example of the capabilities of the INFRA-ICE experimental station, we have presented the UV photochemistry of C$_{11}$H$_{24}$ at low temperature. We have shown that the UV processing of C$_{11}$H$_{24}$ at 14 K promotes the formation of unsaturated hydrocarbon species along with the production of methane. In the near future, realistic cosmic dust analogs and their interaction with ices of astrophysical interest will be investigated with the INFRA-ICE experimental station.

\section{Data availability}
The data that support the findings of this study are available from the corresponding author upon reasonable request.

\acknowledgements
We thank the European Research Council for funding support under Synergy Grant ERC-2013-SyG, G.A. 610256 (NANOCOSMOS). Also, partial support from the Spanish Research Agency (AEI) through grants MAT2017-85089-c2-1R and FIS2016-77578-R is acknowledged. Support from the FotoArt-CM Project (P2018/NMT 4367) through the Program of R$\&$D activities between research groups in Technologies 2013, co-financed by European Structural Funds, is also acknowledged. G.T-C. acknowledges funding from the Comunidad Aut\'onoma de Madrid (PEJD-2018-PRE/IND-9029). G.S. and G.J.E. would like to thank Stephane Lefrançois for valuable discussions on the mechanical details of the optical coupling.

\bibliography{santoro_infra.bib}

\begin{thebibliography}{76}%
\makeatletter
\providecommand \@ifxundefined [1]{%
 \@ifx{#1\undefined}
}%
\providecommand \@ifnum [1]{%
 \ifnum #1\expandafter \@firstoftwo
 \else \expandafter \@secondoftwo
 \fi
}%
\providecommand \@ifx [1]{%
 \ifx #1\expandafter \@firstoftwo
 \else \expandafter \@secondoftwo
 \fi
}%
\providecommand \natexlab [1]{#1}%
\providecommand \enquote  [1]{``#1''}%
\providecommand \bibnamefont  [1]{#1}%
\providecommand \bibfnamefont [1]{#1}%
\providecommand \citenamefont [1]{#1}%
\providecommand \href@noop [0]{\@secondoftwo}%
\providecommand \href [0]{\begingroup \@sanitize@url \@href}%
\providecommand \@href[1]{\@@startlink{#1}\@@href}%
\providecommand \@@href[1]{\endgroup#1\@@endlink}%
\providecommand \@sanitize@url [0]{\catcode `\\12\catcode `\$12\catcode
  `\&12\catcode `\#12\catcode `\^12\catcode `\_12\catcode `\%12\relax}%
\providecommand \@@startlink[1]{}%
\providecommand \@@endlink[0]{}%
\providecommand \url  [0]{\begingroup\@sanitize@url \@url }%
\providecommand \@url [1]{\endgroup\@href {#1}{\urlprefix }}%
\providecommand \urlprefix  [0]{URL }%
\providecommand \Eprint [0]{\href }%
\providecommand \doibase [0]{http://dx.doi.org/}%
\providecommand \selectlanguage [0]{\@gobble}%
\providecommand \bibinfo  [0]{\@secondoftwo}%
\providecommand \bibfield  [0]{\@secondoftwo}%
\providecommand \translation [1]{[#1]}%
\providecommand \BibitemOpen [0]{}%
\providecommand \bibitemStop [0]{}%
\providecommand \bibitemNoStop [0]{.\EOS\space}%
\providecommand \EOS [0]{\spacefactor3000\relax}%
\providecommand \BibitemShut  [1]{\csname bibitem#1\endcsname}%
\let\auto@bib@innerbib\@empty
%</preamble>
\bibitem [{\citenamefont {{Canosa}}\ \emph {et~al.}(1997)\citenamefont
  {{Canosa}}, \citenamefont {{Sims}}, \citenamefont {{Travers}}, \citenamefont
  {{Smith}},\ and\ \citenamefont {{Rowe}}}]{canosa97}%
  \BibitemOpen
  \bibfield  {author} {\bibinfo {author} {\bibfnamefont {A.}~\bibnamefont
  {{Canosa}}}, \bibinfo {author} {\bibfnamefont {I.~R.}\ \bibnamefont
  {{Sims}}}, \bibinfo {author} {\bibfnamefont {D.}~\bibnamefont {{Travers}}},
  \bibinfo {author} {\bibfnamefont {I.~W.~M.}\ \bibnamefont {{Smith}}}, \ and\
  \bibinfo {author} {\bibfnamefont {B.~R.}\ \bibnamefont {{Rowe}}},\ }\bibfield
   {title} {\enquote {\bibinfo {title} {{Reactions of the methylidine radical
  with CH$_4$, C$_2$H$_2$, C$_2$H$_4$, C$_2$H$_6$, and but-1-ene studied
  between 23 and 295K with a CRESU apparatus.}}}\ }\href@noop {} {\bibfield
  {journal} {\bibinfo  {journal} {Astronomy and Astrophysics}\ }\textbf
  {\bibinfo {volume} {323}},\ \bibinfo {pages} {644--651} (\bibinfo {year}
  {1997})}\BibitemShut {NoStop}%
\bibitem [{\citenamefont {Anti{\~{n}}olo}\ \emph {et~al.}(2016)\citenamefont
  {Anti{\~{n}}olo}, \citenamefont {Ag{\'{u}}ndez}, \citenamefont
  {Jim{\'{e}}nez}, \citenamefont {Ballesteros}, \citenamefont {Canosa},
  \citenamefont {Dib}, \citenamefont {Albaladejo},\ and\ \citenamefont
  {Cernicharo}}]{antinolo16}%
  \BibitemOpen
  \bibfield  {author} {\bibinfo {author} {\bibfnamefont {M.}~\bibnamefont
  {Anti{\~{n}}olo}}, \bibinfo {author} {\bibfnamefont {M.}~\bibnamefont
  {Ag{\'{u}}ndez}}, \bibinfo {author} {\bibfnamefont {E.}~\bibnamefont
  {Jim{\'{e}}nez}}, \bibinfo {author} {\bibfnamefont {B.}~\bibnamefont
  {Ballesteros}}, \bibinfo {author} {\bibfnamefont {A.}~\bibnamefont {Canosa}},
  \bibinfo {author} {\bibfnamefont {G.~E.}\ \bibnamefont {Dib}}, \bibinfo
  {author} {\bibfnamefont {J.}~\bibnamefont {Albaladejo}}, \ and\ \bibinfo
  {author} {\bibfnamefont {J.}~\bibnamefont {Cernicharo}},\ }\bibfield  {title}
  {\enquote {\bibinfo {title} {{Reactivity of OH and CH(3)OH Between 22 and 64
  K: Modelling the Gas Phase Production of CH(3)O in Barnard 1B}},}\ }\href
  {\doibase 10.3847/0004-637X/823/1/25} {\bibfield  {journal} {\bibinfo
  {journal} {The Astrophysical Journal}\ }\textbf {\bibinfo {volume} {823}},\
  \bibinfo {pages} {25} (\bibinfo {year} {2016})}\BibitemShut {NoStop}%
\bibitem [{\citenamefont {Potapov}\ \emph {et~al.}(2017)\citenamefont
  {Potapov}, \citenamefont {Canosa}, \citenamefont {Jiménez},\ and\
  \citenamefont {Rowe}}]{potapov17}%
  \BibitemOpen
  \bibfield  {author} {\bibinfo {author} {\bibfnamefont {A.}~\bibnamefont
  {Potapov}}, \bibinfo {author} {\bibfnamefont {A.}~\bibnamefont {Canosa}},
  \bibinfo {author} {\bibfnamefont {E.}~\bibnamefont {Jiménez}}, \ and\
  \bibinfo {author} {\bibfnamefont {B.}~\bibnamefont {Rowe}},\ }\bibfield
  {title} {\enquote {\bibinfo {title} {Uniform supersonic chemical reactors: 30
  years of astrochemical history and future challenges},}\ }\href {\doibase
  10.1002/anie.201611240} {\bibfield  {journal} {\bibinfo  {journal}
  {Angewandte Chemie International Edition}\ }\textbf {\bibinfo {volume}
  {56}},\ \bibinfo {pages} {8618--8640} (\bibinfo {year} {2017})}\BibitemShut
  {NoStop}%
\bibitem [{\citenamefont {{Tanarro}}\ \emph {et~al.}(2018)\citenamefont
  {{Tanarro}}, \citenamefont {{Alem{\'a}n}}, \citenamefont {{de Vicente}},
  \citenamefont {{Gallego}}, \citenamefont {{Pardo}}, \citenamefont
  {{Santoro}}, \citenamefont {{Lauwaet}}, \citenamefont {{Tercero}},
  \citenamefont {{D{\'\i}az-Pulido}}, \citenamefont {{Moreno}}, \citenamefont
  {{Ag{\'u}ndez}}, \citenamefont {{Goicoechea}}, \citenamefont {{Sobrado}},
  \citenamefont {{L{\'o}pez}}, \citenamefont {{Mart{\'\i}nez}}, \citenamefont
  {{Dom{\'e}nech}}, \citenamefont {{Herrero}}, \citenamefont {{Hern{\'a}ndez}},
  \citenamefont {{Pel{\'a}ez}}, \citenamefont {{L{\'o}pez-P{\'e}rez}},
  \citenamefont {{G{\'o}mez-Gonz{\'a}lez}}, \citenamefont {{Alonso}},
  \citenamefont {{Jim{\'e}nez}}, \citenamefont {{Teyssier}}, \citenamefont
  {{Makasheva}}, \citenamefont {{Castellanos}}, \citenamefont {{Joblin}},
  \citenamefont {{Mart{\'\i}n-Gago}},\ and\ \citenamefont
  {{Cernicharo}}}]{tanarro18}%
  \BibitemOpen
  \bibfield  {author} {\bibinfo {author} {\bibfnamefont {I.}~\bibnamefont
  {{Tanarro}}}, \bibinfo {author} {\bibfnamefont {B.}~\bibnamefont
  {{Alem{\'a}n}}}, \bibinfo {author} {\bibfnamefont {P.}~\bibnamefont {{de
  Vicente}}}, \bibinfo {author} {\bibfnamefont {J.~D.}\ \bibnamefont
  {{Gallego}}}, \bibinfo {author} {\bibfnamefont {J.~R.}\ \bibnamefont
  {{Pardo}}}, \bibinfo {author} {\bibfnamefont {G.}~\bibnamefont {{Santoro}}},
  \bibinfo {author} {\bibfnamefont {K.}~\bibnamefont {{Lauwaet}}}, \bibinfo
  {author} {\bibfnamefont {F.}~\bibnamefont {{Tercero}}}, \bibinfo {author}
  {\bibfnamefont {A.}~\bibnamefont {{D{\'\i}az-Pulido}}}, \bibinfo {author}
  {\bibfnamefont {E.}~\bibnamefont {{Moreno}}}, \bibinfo {author}
  {\bibfnamefont {M.}~\bibnamefont {{Ag{\'u}ndez}}}, \bibinfo {author}
  {\bibfnamefont {J.~R.}\ \bibnamefont {{Goicoechea}}}, \bibinfo {author}
  {\bibfnamefont {J.~M.}\ \bibnamefont {{Sobrado}}}, \bibinfo {author}
  {\bibfnamefont {J.~A.}\ \bibnamefont {{L{\'o}pez}}}, \bibinfo {author}
  {\bibfnamefont {L.}~\bibnamefont {{Mart{\'\i}nez}}}, \bibinfo {author}
  {\bibfnamefont {J.~L.}\ \bibnamefont {{Dom{\'e}nech}}}, \bibinfo {author}
  {\bibfnamefont {V.~J.}\ \bibnamefont {{Herrero}}}, \bibinfo {author}
  {\bibfnamefont {J.~M.}\ \bibnamefont {{Hern{\'a}ndez}}}, \bibinfo {author}
  {\bibfnamefont {R.~J.}\ \bibnamefont {{Pel{\'a}ez}}}, \bibinfo {author}
  {\bibfnamefont {J.~A.}\ \bibnamefont {{L{\'o}pez-P{\'e}rez}}}, \bibinfo
  {author} {\bibfnamefont {J.}~\bibnamefont {{G{\'o}mez-Gonz{\'a}lez}}},
  \bibinfo {author} {\bibfnamefont {J.~L.}\ \bibnamefont {{Alonso}}}, \bibinfo
  {author} {\bibfnamefont {E.}~\bibnamefont {{Jim{\'e}nez}}}, \bibinfo {author}
  {\bibfnamefont {D.}~\bibnamefont {{Teyssier}}}, \bibinfo {author}
  {\bibfnamefont {K.}~\bibnamefont {{Makasheva}}}, \bibinfo {author}
  {\bibfnamefont {M.}~\bibnamefont {{Castellanos}}}, \bibinfo {author}
  {\bibfnamefont {C.}~\bibnamefont {{Joblin}}}, \bibinfo {author}
  {\bibfnamefont {J.~A.}\ \bibnamefont {{Mart{\'\i}n-Gago}}}, \ and\ \bibinfo
  {author} {\bibfnamefont {J.}~\bibnamefont {{Cernicharo}}},\ }\bibfield
  {title} {\enquote {\bibinfo {title} {{Using radio astronomical receivers for
  molecular spectroscopic characterization in astrochemical laboratory
  simulations: A proof of concept}},}\ }\href {\doibase
  10.1051/0004-6361/201730969} {\bibfield  {journal} {\bibinfo  {journal}
  {Astronomy \& Astrophysics}\ }\textbf {\bibinfo {volume} {609}},\ \bibinfo
  {eid} {A15} (\bibinfo {year} {2018})}\BibitemShut {NoStop}%
\bibitem [{\citenamefont {{Cernicharo}}\ \emph {et~al.}(2019)\citenamefont
  {{Cernicharo}}, \citenamefont {{Gallego}}, \citenamefont
  {{L{\'o}pez-P{\'e}rez}}, \citenamefont {{Tercero}}, \citenamefont
  {{Tanarro}}, \citenamefont {{Beltr{\'a}n}}, \citenamefont {{de Vicente}},
  \citenamefont {{Lauwaet}}, \citenamefont {{Alem{\'a}n}}, \citenamefont
  {{Moreno}}, \citenamefont {{Herrero}}, \citenamefont {{Dom{\'e}nech}},
  \citenamefont {{Ram{\'\i}rez}}, \citenamefont {{Berm{\'u}dez}}, \citenamefont
  {{Pel{\'a}ez}}, \citenamefont {{Patino-Esteban}}, \citenamefont
  {{L{\'o}pez-Fern{\'a}ndez}}, \citenamefont {{Garc{\'\i}a-{\'A}lvaro}},
  \citenamefont {{Garc{\'\i}a-Carre{\~n}o}}, \citenamefont {{Cabezas}},
  \citenamefont {{Malo}}, \citenamefont {{Amils}}, \citenamefont {{Sobrado}},
  \citenamefont {{Diez-Gonz{\'a}lez}}, \citenamefont {{Hernand{\'e}z}},
  \citenamefont {{Tercero}}, \citenamefont {{Santoro}}, \citenamefont
  {{Mart{\'\i}nez}}, \citenamefont {{Castellanos}}, \citenamefont {{Vaquero
  Jim{\'e}nez}}, \citenamefont {{Pardo}}, \citenamefont {{Barbas}},
  \citenamefont {{L{\'o}pez-Fern{\'a}ndez}}, \citenamefont {{Aja}},
  \citenamefont {{Leuther}},\ and\ \citenamefont
  {{Mart{\'\i}n-Gago}}}]{cernicharo19}%
  \BibitemOpen
  \bibfield  {author} {\bibinfo {author} {\bibfnamefont {J.}~\bibnamefont
  {{Cernicharo}}}, \bibinfo {author} {\bibfnamefont {J.~D.}\ \bibnamefont
  {{Gallego}}}, \bibinfo {author} {\bibfnamefont {J.~A.}\ \bibnamefont
  {{L{\'o}pez-P{\'e}rez}}}, \bibinfo {author} {\bibfnamefont {F.}~\bibnamefont
  {{Tercero}}}, \bibinfo {author} {\bibfnamefont {I.}~\bibnamefont
  {{Tanarro}}}, \bibinfo {author} {\bibfnamefont {F.}~\bibnamefont
  {{Beltr{\'a}n}}}, \bibinfo {author} {\bibfnamefont {P.}~\bibnamefont {{de
  Vicente}}}, \bibinfo {author} {\bibfnamefont {K.}~\bibnamefont {{Lauwaet}}},
  \bibinfo {author} {\bibfnamefont {B.}~\bibnamefont {{Alem{\'a}n}}}, \bibinfo
  {author} {\bibfnamefont {E.}~\bibnamefont {{Moreno}}}, \bibinfo {author}
  {\bibfnamefont {V.~J.}\ \bibnamefont {{Herrero}}}, \bibinfo {author}
  {\bibfnamefont {J.~L.}\ \bibnamefont {{Dom{\'e}nech}}}, \bibinfo {author}
  {\bibfnamefont {S.~I.}\ \bibnamefont {{Ram{\'\i}rez}}}, \bibinfo {author}
  {\bibfnamefont {C.}~\bibnamefont {{Berm{\'u}dez}}}, \bibinfo {author}
  {\bibfnamefont {R.~J.}\ \bibnamefont {{Pel{\'a}ez}}}, \bibinfo {author}
  {\bibfnamefont {M.}~\bibnamefont {{Patino-Esteban}}}, \bibinfo {author}
  {\bibfnamefont {I.}~\bibnamefont {{L{\'o}pez-Fern{\'a}ndez}}}, \bibinfo
  {author} {\bibfnamefont {S.}~\bibnamefont {{Garc{\'\i}a-{\'A}lvaro}}},
  \bibinfo {author} {\bibfnamefont {P.}~\bibnamefont
  {{Garc{\'\i}a-Carre{\~n}o}}}, \bibinfo {author} {\bibfnamefont
  {C.}~\bibnamefont {{Cabezas}}}, \bibinfo {author} {\bibfnamefont
  {I.}~\bibnamefont {{Malo}}}, \bibinfo {author} {\bibfnamefont
  {R.}~\bibnamefont {{Amils}}}, \bibinfo {author} {\bibfnamefont
  {J.}~\bibnamefont {{Sobrado}}}, \bibinfo {author} {\bibfnamefont
  {C.}~\bibnamefont {{Diez-Gonz{\'a}lez}}}, \bibinfo {author} {\bibfnamefont
  {J.~M.}\ \bibnamefont {{Hernand{\'e}z}}}, \bibinfo {author} {\bibfnamefont
  {B.}~\bibnamefont {{Tercero}}}, \bibinfo {author} {\bibfnamefont
  {G.}~\bibnamefont {{Santoro}}}, \bibinfo {author} {\bibfnamefont
  {L.}~\bibnamefont {{Mart{\'\i}nez}}}, \bibinfo {author} {\bibfnamefont
  {M.}~\bibnamefont {{Castellanos}}}, \bibinfo {author} {\bibfnamefont
  {B.}~\bibnamefont {{Vaquero Jim{\'e}nez}}}, \bibinfo {author} {\bibfnamefont
  {J.~R.}\ \bibnamefont {{Pardo}}}, \bibinfo {author} {\bibfnamefont
  {L.}~\bibnamefont {{Barbas}}}, \bibinfo {author} {\bibfnamefont {J.~A.}\
  \bibnamefont {{L{\'o}pez-Fern{\'a}ndez}}}, \bibinfo {author} {\bibfnamefont
  {B.}~\bibnamefont {{Aja}}}, \bibinfo {author} {\bibfnamefont
  {A.}~\bibnamefont {{Leuther}}}, \ and\ \bibinfo {author} {\bibfnamefont
  {J.~A.}\ \bibnamefont {{Mart{\'\i}n-Gago}}},\ }\bibfield  {title} {\enquote
  {\bibinfo {title} {{Broad-band high-resolution rotational spectroscopy for
  laboratory astrophysics}},}\ }\href {\doibase 10.1051/0004-6361/201935197}
  {\bibfield  {journal} {\bibinfo  {journal} {Astronomy \& Astrophysics}\
  }\textbf {\bibinfo {volume} {626}},\ \bibinfo {eid} {A34} (\bibinfo {year}
  {2019})}\BibitemShut {NoStop}%
\bibitem [{\citenamefont {Sobrado}, \citenamefont {Martín-Soler},\ and\
  \citenamefont {Martín-Gago}(2014)}]{sobrado14}%
  \BibitemOpen
  \bibfield  {author} {\bibinfo {author} {\bibfnamefont {J.~M.}\ \bibnamefont
  {Sobrado}}, \bibinfo {author} {\bibfnamefont {J.}~\bibnamefont
  {Martín-Soler}}, \ and\ \bibinfo {author} {\bibfnamefont {J.~A.}\
  \bibnamefont {Martín-Gago}},\ }\bibfield  {title} {\enquote {\bibinfo
  {title} {Mimicking mars: A vacuum simulation chamber for testing
  environmental instrumentation for mars exploration},}\ }\href {\doibase
  10.1063/1.4868592} {\bibfield  {journal} {\bibinfo  {journal} {Review of
  Scientific Instruments}\ }\textbf {\bibinfo {volume} {85}},\ \bibinfo {pages}
  {035111} (\bibinfo {year} {2014})}\BibitemShut {NoStop}%
\bibitem [{\citenamefont {Sobrado}, \citenamefont {Martín-Soler},\ and\
  \citenamefont {Martín-Gago}(2015)}]{sobrado15}%
  \BibitemOpen
  \bibfield  {author} {\bibinfo {author} {\bibfnamefont {J.~M.}\ \bibnamefont
  {Sobrado}}, \bibinfo {author} {\bibfnamefont {J.}~\bibnamefont
  {Martín-Soler}}, \ and\ \bibinfo {author} {\bibfnamefont {J.~A.}\
  \bibnamefont {Martín-Gago}},\ }\bibfield  {title} {\enquote {\bibinfo
  {title} {Mimicking martian dust: An in-vacuum dust deposition system for
  testing the ultraviolet sensors on the curiosity rover},}\ }\href {\doibase
  10.1063/1.4932937} {\bibfield  {journal} {\bibinfo  {journal} {Review of
  Scientific Instruments}\ }\textbf {\bibinfo {volume} {86}},\ \bibinfo {pages}
  {105113} (\bibinfo {year} {2015})}\BibitemShut {NoStop}%
\bibitem [{\citenamefont {Romanini}\ \emph {et~al.}(1999)\citenamefont
  {Romanini}, \citenamefont {Biennier}, \citenamefont {Salama}, \citenamefont
  {Kachanov}, \citenamefont {Allamandola},\ and\ \citenamefont
  {Stoeckel}}]{romanini99}%
  \BibitemOpen
  \bibfield  {author} {\bibinfo {author} {\bibfnamefont {D.}~\bibnamefont
  {Romanini}}, \bibinfo {author} {\bibfnamefont {L.}~\bibnamefont {Biennier}},
  \bibinfo {author} {\bibfnamefont {F.}~\bibnamefont {Salama}}, \bibinfo
  {author} {\bibfnamefont {A.}~\bibnamefont {Kachanov}}, \bibinfo {author}
  {\bibfnamefont {L.}~\bibnamefont {Allamandola}}, \ and\ \bibinfo {author}
  {\bibfnamefont {F.}~\bibnamefont {Stoeckel}},\ }\bibfield  {title} {\enquote
  {\bibinfo {title} {Jet-discharge cavity ring-down spectroscopy of ionized
  polycyclic aromatic hydrocarbons: progress in testing the pah hypothesis for
  the diffuse interstellar band problem},}\ }\href {\doibase
  https://doi.org/10.1016/S0009-2614(99)00210-9} {\bibfield  {journal}
  {\bibinfo  {journal} {Chemical Physics Letters}\ }\textbf {\bibinfo {volume}
  {303}},\ \bibinfo {pages} {165 -- 170} (\bibinfo {year} {1999})}\BibitemShut
  {NoStop}%
\bibitem [{\citenamefont {{Br{\'e}chignac}}\ and\ \citenamefont
  {{Pino}}(1999)}]{brechignac99}%
  \BibitemOpen
  \bibfield  {author} {\bibinfo {author} {\bibfnamefont {P.}~\bibnamefont
  {{Br{\'e}chignac}}}\ and\ \bibinfo {author} {\bibfnamefont {T.}~\bibnamefont
  {{Pino}}},\ }\bibfield  {title} {\enquote {\bibinfo {title} {{Electronic
  spectra of cold gas phase PAH cations: Towards the identification of the
  Diffuse Interstellar Bands carriers}},}\ }\href@noop {} {\bibfield  {journal}
  {\bibinfo  {journal} {Astronomy and astrophysics}\ }\textbf {\bibinfo
  {volume} {343}},\ \bibinfo {pages} {L49--L52} (\bibinfo {year}
  {1999})}\BibitemShut {NoStop}%
\bibitem [{\citenamefont {{Joblin, C.}}\ \emph {et~al.}(2002)\citenamefont
  {{Joblin, C.}}, \citenamefont {{Pech, C.}}, \citenamefont {{Armengaud, M.}},
  \citenamefont {{Frabel, P.}},\ and\ \citenamefont {{Boissel,
  P.}}}]{joblin02}%
  \BibitemOpen
  \bibfield  {author} {\bibinfo {author} {\bibnamefont {{Joblin, C.}}},
  \bibinfo {author} {\bibnamefont {{Pech, C.}}}, \bibinfo {author}
  {\bibnamefont {{Armengaud, M.}}}, \bibinfo {author} {\bibnamefont {{Frabel,
  P.}}}, \ and\ \bibinfo {author} {\bibnamefont {{Boissel, P.}}},\ }\bibfield
  {title} {\enquote {\bibinfo {title} {A piece of interstellar medium in the
  laboratory: the pirenea experiment},}\ }\href {\doibase 10.1051/eas:2002061}
  {\bibfield  {journal} {\bibinfo  {journal} {EAS Publications Series}\
  }\textbf {\bibinfo {volume} {4}},\ \bibinfo {pages} {73} (\bibinfo {year}
  {2002})}\BibitemShut {NoStop}%
\bibitem [{\citenamefont {Biennier}\ \emph {et~al.}(2003)\citenamefont
  {Biennier}, \citenamefont {Salama}, \citenamefont {Allamandola},\ and\
  \citenamefont {Scherer}}]{biennier03}%
  \BibitemOpen
  \bibfield  {author} {\bibinfo {author} {\bibfnamefont {L.}~\bibnamefont
  {Biennier}}, \bibinfo {author} {\bibfnamefont {F.}~\bibnamefont {Salama}},
  \bibinfo {author} {\bibfnamefont {L.~J.}\ \bibnamefont {Allamandola}}, \ and\
  \bibinfo {author} {\bibfnamefont {J.~J.}\ \bibnamefont {Scherer}},\
  }\bibfield  {title} {\enquote {\bibinfo {title} {Pulsed discharge nozzle
  cavity ringdown spectroscopy of cold polycyclic aromatic hydrocarbon ions},}\
  }\href {\doibase 10.1063/1.1564044} {\bibfield  {journal} {\bibinfo
  {journal} {The Journal of Chemical Physics}\ }\textbf {\bibinfo {volume}
  {118}},\ \bibinfo {pages} {7863--7872} (\bibinfo {year} {2003})}\BibitemShut
  {NoStop}%
\bibitem [{\citenamefont {{Kaiser}}\ and\ \citenamefont
  {{Osamura}}(2005)}]{kaiser05a}%
  \BibitemOpen
  \bibfield  {author} {\bibinfo {author} {\bibfnamefont {R.~I.}\ \bibnamefont
  {{Kaiser}}}\ and\ \bibinfo {author} {\bibfnamefont {Y.}~\bibnamefont
  {{Osamura}}},\ }\bibfield  {title} {\enquote {\bibinfo {title} {{Infrared
  spectroscopic studies of hydrogenated silicon clusters. Guiding the search
  for Si$_{2}$H$_{x}$ species in the Circumstellar Envelope of IRC+10216}},}\
  }\href {\doibase 10.1051/0004-6361:20040305} {\bibfield  {journal} {\bibinfo
  {journal} {Astronomy \& Astrophysics}\ }\textbf {\bibinfo {volume} {432}},\
  \bibinfo {pages} {559--566} (\bibinfo {year} {2005})}\BibitemShut {NoStop}%
\bibitem [{\citenamefont {Useli-Bacchitta}\ \emph {et~al.}(2010)\citenamefont
  {Useli-Bacchitta}, \citenamefont {Bonnamy}, \citenamefont {Mulas},
  \citenamefont {Malloci}, \citenamefont {Toublanc},\ and\ \citenamefont
  {Joblin}}]{useli10}%
  \BibitemOpen
  \bibfield  {author} {\bibinfo {author} {\bibfnamefont {F.}~\bibnamefont
  {Useli-Bacchitta}}, \bibinfo {author} {\bibfnamefont {A.}~\bibnamefont
  {Bonnamy}}, \bibinfo {author} {\bibfnamefont {G.}~\bibnamefont {Mulas}},
  \bibinfo {author} {\bibfnamefont {G.}~\bibnamefont {Malloci}}, \bibinfo
  {author} {\bibfnamefont {D.}~\bibnamefont {Toublanc}}, \ and\ \bibinfo
  {author} {\bibfnamefont {C.}~\bibnamefont {Joblin}},\ }\bibfield  {title}
  {\enquote {\bibinfo {title} {Visible photodissociation spectroscopy of pah
  cations and derivatives in the pirenea experiment},}\ }\href {\doibase
  https://doi.org/10.1016/j.chemphys.2010.03.012} {\bibfield  {journal}
  {\bibinfo  {journal} {Chemical Physics}\ }\textbf {\bibinfo {volume} {371}},\
  \bibinfo {pages} {16 -- 23} (\bibinfo {year} {2010})}\BibitemShut {NoStop}%
\bibitem [{\citenamefont {Asvany}\ \emph {et~al.}(2010)\citenamefont {Asvany},
  \citenamefont {Bielau}, \citenamefont {Moratschke}, \citenamefont {Krause},\
  and\ \citenamefont {Schlemmer}}]{asvany10}%
  \BibitemOpen
  \bibfield  {author} {\bibinfo {author} {\bibfnamefont {O.}~\bibnamefont
  {Asvany}}, \bibinfo {author} {\bibfnamefont {F.}~\bibnamefont {Bielau}},
  \bibinfo {author} {\bibfnamefont {D.}~\bibnamefont {Moratschke}}, \bibinfo
  {author} {\bibfnamefont {J.}~\bibnamefont {Krause}}, \ and\ \bibinfo {author}
  {\bibfnamefont {S.}~\bibnamefont {Schlemmer}},\ }\bibfield  {title} {\enquote
  {\bibinfo {title} {Note: {New} design of a cryogenic linear radio frequency
  multipole trap},}\ }\href {\doibase 10.1063/1.3460265} {\bibfield  {journal}
  {\bibinfo  {journal} {Review of Scientific Instruments}\ }\textbf {\bibinfo
  {volume} {81}},\ \bibinfo {pages} {076102} (\bibinfo {year}
  {2010})}\BibitemShut {NoStop}%
\bibitem [{\citenamefont {Campbell}\ and\ \citenamefont
  {Maier}(2017)}]{campbell17}%
  \BibitemOpen
  \bibfield  {author} {\bibinfo {author} {\bibfnamefont {E.~K.}\ \bibnamefont
  {Campbell}}\ and\ \bibinfo {author} {\bibfnamefont {J.~P.}\ \bibnamefont
  {Maier}},\ }\bibfield  {title} {\enquote {\bibinfo {title} {Perspective: C60+
  and laboratory spectroscopy related to diffuse interstellar bands},}\ }\href
  {\doibase 10.1063/1.4980119} {\bibfield  {journal} {\bibinfo  {journal} {The
  Journal of Chemical Physics}\ }\textbf {\bibinfo {volume} {146}},\ \bibinfo
  {pages} {160901} (\bibinfo {year} {2017})}\BibitemShut {NoStop}%
\bibitem [{\citenamefont {Dom{\'{e}}nech}, \citenamefont {Schlemmer},\ and\
  \citenamefont {Asvany}(2018)}]{domenech18}%
  \BibitemOpen
  \bibfield  {author} {\bibinfo {author} {\bibfnamefont {J.~L.}\ \bibnamefont
  {Dom{\'{e}}nech}}, \bibinfo {author} {\bibfnamefont {S.}~\bibnamefont
  {Schlemmer}}, \ and\ \bibinfo {author} {\bibfnamefont {O.}~\bibnamefont
  {Asvany}},\ }\bibfield  {title} {\enquote {\bibinfo {title} {Accurate
  rotational rest frequencies for ammonium ion isotopologues},}\ }\href
  {https://doi.org/10.3847\%2F1538-4357\%2Faadf83} {\bibfield  {journal}
  {\bibinfo  {journal} {The Astrophysical Journal}\ }\textbf {\bibinfo {volume}
  {866}},\ \bibinfo {pages} {158} (\bibinfo {year} {2018})}\BibitemShut
  {NoStop}%
\bibitem [{\citenamefont {Fern{\'{a}}ndez}\ \emph {et~al.}(2019)\citenamefont
  {Fern{\'{a}}ndez}, \citenamefont {Tejeda}, \citenamefont {Carvajal},\ and\
  \citenamefont {Senent}}]{fernandez19}%
  \BibitemOpen
  \bibfield  {author} {\bibinfo {author} {\bibfnamefont {J.~M.}\ \bibnamefont
  {Fern{\'{a}}ndez}}, \bibinfo {author} {\bibfnamefont {G.}~\bibnamefont
  {Tejeda}}, \bibinfo {author} {\bibfnamefont {M.}~\bibnamefont {Carvajal}}, \
  and\ \bibinfo {author} {\bibfnamefont {M.~L.}\ \bibnamefont {Senent}},\
  }\bibfield  {title} {\enquote {\bibinfo {title} {New spectral
  characterization of dimethyl ether isotopologues {CH$_3$OCH$_3$} and
  {$^{13}$CH$_3$OCH$_3$} in the {THz} region},}\ }\href {\doibase
  10.3847/1538-4365/ab041e} {\bibfield  {journal} {\bibinfo  {journal} {The
  Astrophysical Journal Supplement Series}\ }\textbf {\bibinfo {volume}
  {241}},\ \bibinfo {pages} {13} (\bibinfo {year} {2019})}\BibitemShut
  {NoStop}%
\bibitem [{\citenamefont {Jäger}\ \emph {et~al.}(2009)\citenamefont {Jäger},
  \citenamefont {Huisken}, \citenamefont {Mutschke}, \citenamefont {Jansa},\
  and\ \citenamefont {Henning}}]{jager09}%
  \BibitemOpen
  \bibfield  {author} {\bibinfo {author} {\bibfnamefont {C.}~\bibnamefont
  {Jäger}}, \bibinfo {author} {\bibfnamefont {F.}~\bibnamefont {Huisken}},
  \bibinfo {author} {\bibfnamefont {H.}~\bibnamefont {Mutschke}}, \bibinfo
  {author} {\bibfnamefont {I.~L.}\ \bibnamefont {Jansa}}, \ and\ \bibinfo
  {author} {\bibfnamefont {T.}~\bibnamefont {Henning}},\ }\bibfield  {title}
  {\enquote {\bibinfo {title} {Formation of polycyclic aromatic hydrocarbons
  and carbonaceous solids in gas-phase condensation experiments},}\ }\href
  {\doibase 10.1088/0004-637x/696/1/706} {\bibfield  {journal} {\bibinfo
  {journal} {The Astrophysical Journal}\ }\textbf {\bibinfo {volume} {696}},\
  \bibinfo {pages} {706--712} (\bibinfo {year} {2009})}\BibitemShut {NoStop}%
\bibitem [{\citenamefont {{Contreras}}\ and\ \citenamefont
  {{Salama}}(2013)}]{contreras13}%
  \BibitemOpen
  \bibfield  {author} {\bibinfo {author} {\bibfnamefont {C.~S.}\ \bibnamefont
  {{Contreras}}}\ and\ \bibinfo {author} {\bibfnamefont {F.}~\bibnamefont
  {{Salama}}},\ }\bibfield  {title} {\enquote {\bibinfo {title} {{Laboratory
  Investigations of Polycyclic Aromatic Hydrocarbon Formation and Destruction
  in the Circumstellar Outflows of Carbon Stars}},}\ }\href {\doibase
  10.1088/0067-0049/208/1/6} {\bibfield  {journal} {\bibinfo  {journal} {The
  Astrophysical Journal Supplement Series}\ }\textbf {\bibinfo {volume}
  {208}},\ \bibinfo {eid} {6} (\bibinfo {year} {2013})}\BibitemShut {NoStop}%
\bibitem [{\citenamefont {Mart\'i{}nez}\ \emph {et~al.}(2020)\citenamefont
  {Mart\'i{}nez}, \citenamefont {Santoro}, \citenamefont {Merino},
  \citenamefont {Accolla}, \citenamefont {Lauwaet}, \citenamefont {Sobrado},
  \citenamefont {Sabbah}, \citenamefont {Pelaez}, \citenamefont {Herrero},
  \citenamefont {Tanarro}, \citenamefont {Agúndez}, \citenamefont
  {Martín-Jimenez}, \citenamefont {Otero}, \citenamefont {Ellis},
  \citenamefont {Joblin}, \citenamefont {Cernicharo},\ and\ \citenamefont
  {Martín-Gago}}]{martinez20}%
  \BibitemOpen
  \bibfield  {author} {\bibinfo {author} {\bibfnamefont {L.}~\bibnamefont
  {Mart\'i{}nez}}, \bibinfo {author} {\bibfnamefont {G.}~\bibnamefont
  {Santoro}}, \bibinfo {author} {\bibfnamefont {P.}~\bibnamefont {Merino}},
  \bibinfo {author} {\bibfnamefont {M.}~\bibnamefont {Accolla}}, \bibinfo
  {author} {\bibfnamefont {K.}~\bibnamefont {Lauwaet}}, \bibinfo {author}
  {\bibfnamefont {J.}~\bibnamefont {Sobrado}}, \bibinfo {author} {\bibfnamefont
  {H.}~\bibnamefont {Sabbah}}, \bibinfo {author} {\bibfnamefont {R.~J.}\
  \bibnamefont {Pelaez}}, \bibinfo {author} {\bibfnamefont {V.~J.}\
  \bibnamefont {Herrero}}, \bibinfo {author} {\bibfnamefont {I.}~\bibnamefont
  {Tanarro}}, \bibinfo {author} {\bibfnamefont {M.}~\bibnamefont {Agúndez}},
  \bibinfo {author} {\bibfnamefont {A.}~\bibnamefont {Martín-Jimenez}},
  \bibinfo {author} {\bibfnamefont {R.}~\bibnamefont {Otero}}, \bibinfo
  {author} {\bibfnamefont {G.~J.}\ \bibnamefont {Ellis}}, \bibinfo {author}
  {\bibfnamefont {C.}~\bibnamefont {Joblin}}, \bibinfo {author} {\bibfnamefont
  {J.}~\bibnamefont {Cernicharo}}, \ and\ \bibinfo {author} {\bibfnamefont
  {J.~A.}\ \bibnamefont {Martín-Gago}},\ }\bibfield  {title} {\enquote
  {\bibinfo {title} {Prevalence of non-aromatic carbonaceous molecules in the
  inner regions of circumstellar envelopes},}\ }\href {\doibase
  10.1038/s41550-019-0899-4} {\bibfield  {journal} {\bibinfo  {journal} {Nature
  Astronomy}\ }\textbf {\bibinfo {volume} {4}},\ \bibinfo {pages} {97--105}
  (\bibinfo {year} {2020})}\BibitemShut {NoStop}%
\bibitem [{\citenamefont {{Santoro}}\ \emph {et~al.}(2020)\citenamefont
  {{Santoro}}, \citenamefont {{Mart{\'\i}nez}}, \citenamefont {{Lauwaet}},
  \citenamefont {{Accolla}}, \citenamefont {{Tajuelo-Castilla}}, \citenamefont
  {{Merino}}, \citenamefont {{Sobrado}}, \citenamefont {{Pel{\'a}ez}},
  \citenamefont {{Herrero}}, \citenamefont {{Tanarro}}, \citenamefont
  {{Mayoral}}, \citenamefont {{Ag{\'u}ndez}}, \citenamefont {{Sabbah}},
  \citenamefont {{Joblin}}, \citenamefont {{Cernicharo}},\ and\ \citenamefont
  {{Mart{\'\i}n-Gago}}}]{santoro20}%
  \BibitemOpen
  \bibfield  {author} {\bibinfo {author} {\bibfnamefont {G.}~\bibnamefont
  {{Santoro}}}, \bibinfo {author} {\bibfnamefont {L.}~\bibnamefont
  {{Mart{\'\i}nez}}}, \bibinfo {author} {\bibfnamefont {K.}~\bibnamefont
  {{Lauwaet}}}, \bibinfo {author} {\bibfnamefont {M.}~\bibnamefont
  {{Accolla}}}, \bibinfo {author} {\bibfnamefont {G.}~\bibnamefont
  {{Tajuelo-Castilla}}}, \bibinfo {author} {\bibfnamefont {P.}~\bibnamefont
  {{Merino}}}, \bibinfo {author} {\bibfnamefont {J.~M.}\ \bibnamefont
  {{Sobrado}}}, \bibinfo {author} {\bibfnamefont {R.~J.}\ \bibnamefont
  {{Pel{\'a}ez}}}, \bibinfo {author} {\bibfnamefont {V.~J.}\ \bibnamefont
  {{Herrero}}}, \bibinfo {author} {\bibfnamefont {I.}~\bibnamefont
  {{Tanarro}}}, \bibinfo {author} {\bibfnamefont {{\'A}.}~\bibnamefont
  {{Mayoral}}}, \bibinfo {author} {\bibfnamefont {M.}~\bibnamefont
  {{Ag{\'u}ndez}}}, \bibinfo {author} {\bibfnamefont {H.}~\bibnamefont
  {{Sabbah}}}, \bibinfo {author} {\bibfnamefont {C.}~\bibnamefont {{Joblin}}},
  \bibinfo {author} {\bibfnamefont {J.}~\bibnamefont {{Cernicharo}}}, \ and\
  \bibinfo {author} {\bibfnamefont {J.~{\'A}.}\ \bibnamefont
  {{Mart{\'\i}n-Gago}}},\ }\bibfield  {title} {\enquote {\bibinfo {title} {{The
  Chemistry of Cosmic Dust Analogs from C, C$_{2}$, and C$_{2}$H$_{2}$ in
  C-rich Circumstellar Envelopes}},}\ }\href {\doibase
  10.3847/1538-4357/ab9086} {\bibfield  {journal} {\bibinfo  {journal} {The
  Astrophysical Journal}\ }\textbf {\bibinfo {volume} {895}},\ \bibinfo {eid}
  {97} (\bibinfo {year} {2020})}\BibitemShut {NoStop}%
\bibitem [{\citenamefont {{Roser}}\ \emph {et~al.}(2001)\citenamefont
  {{Roser}}, \citenamefont {{Vidali}}, \citenamefont {{Manico}},\ and\
  \citenamefont {{Pirronello}}}]{roser01}%
  \BibitemOpen
  \bibfield  {author} {\bibinfo {author} {\bibfnamefont {J.~E.}\ \bibnamefont
  {{Roser}}}, \bibinfo {author} {\bibfnamefont {G.}~\bibnamefont {{Vidali}}},
  \bibinfo {author} {\bibfnamefont {G.}~\bibnamefont {{Manico}}}, \ and\
  \bibinfo {author} {\bibfnamefont {V.}~\bibnamefont {{Pirronello}}},\
  }\bibfield  {title} {\enquote {\bibinfo {title} {{Formation of Carbon Dioxide
  by Surface Reactions on Ices in the Interstellar Medium}},}\ }\href {\doibase
  10.1086/321732} {\bibfield  {journal} {\bibinfo  {journal} {The Astrophysical
  Journal}\ }\textbf {\bibinfo {volume} {555}},\ \bibinfo {pages} {L61--L64}
  (\bibinfo {year} {2001})}\BibitemShut {NoStop}%
\bibitem [{\citenamefont {Mennella}\ \emph {et~al.}(2006)\citenamefont
  {Mennella}, \citenamefont {Baratta}, \citenamefont {Palumbo},\ and\
  \citenamefont {Bergin}}]{mennella06}%
  \BibitemOpen
  \bibfield  {author} {\bibinfo {author} {\bibfnamefont {V.}~\bibnamefont
  {Mennella}}, \bibinfo {author} {\bibfnamefont {G.~A.}\ \bibnamefont
  {Baratta}}, \bibinfo {author} {\bibfnamefont {M.~E.}\ \bibnamefont
  {Palumbo}}, \ and\ \bibinfo {author} {\bibfnamefont {E.~A.}\ \bibnamefont
  {Bergin}},\ }\bibfield  {title} {\enquote {\bibinfo {title} {{Synthesis of CO
  and CO2Molecules by UV Irradiation of Water Ice–covered Hydrogenated Carbon
  Grains}},}\ }\href {\doibase 10.1086/502965} {\bibfield  {journal} {\bibinfo
  {journal} {The Astrophysical Journal}\ }\textbf {\bibinfo {volume} {643}},\
  \bibinfo {pages} {923--931} (\bibinfo {year} {2006})}\BibitemShut {NoStop}%
\bibitem [{\citenamefont {Oba}\ \emph {et~al.}(2009)\citenamefont {Oba},
  \citenamefont {Miyauchi}, \citenamefont {Hidaka}, \citenamefont {Chigai},
  \citenamefont {Watanabe},\ and\ \citenamefont {Kouchi}}]{oba09}%
  \BibitemOpen
  \bibfield  {author} {\bibinfo {author} {\bibfnamefont {Y.}~\bibnamefont
  {Oba}}, \bibinfo {author} {\bibfnamefont {N.}~\bibnamefont {Miyauchi}},
  \bibinfo {author} {\bibfnamefont {H.}~\bibnamefont {Hidaka}}, \bibinfo
  {author} {\bibfnamefont {T.}~\bibnamefont {Chigai}}, \bibinfo {author}
  {\bibfnamefont {N.}~\bibnamefont {Watanabe}}, \ and\ \bibinfo {author}
  {\bibfnamefont {A.}~\bibnamefont {Kouchi}},\ }\bibfield  {title} {\enquote
  {\bibinfo {title} {{FORMATION OF COMPACT AMORPHOUS H2O ICE BY CODEPOSITION OF
  HYDROGEN ATOMS WITH OXYGEN MOLECULES ON GRAIN SURFACES}},}\ }\href {\doibase
  10.1088/0004-637x/701/1/464} {\bibfield  {journal} {\bibinfo  {journal} {The
  Astrophysical Journal}\ }\textbf {\bibinfo {volume} {701}},\ \bibinfo {pages}
  {464--470} (\bibinfo {year} {2009})}\BibitemShut {NoStop}%
\bibitem [{\citenamefont {Palumbo}\ \emph {et~al.}(2010)\citenamefont
  {Palumbo}, \citenamefont {Baratta}, \citenamefont {Leto},\ and\ \citenamefont
  {Strazzulla}}]{palumbo10}%
  \BibitemOpen
  \bibfield  {author} {\bibinfo {author} {\bibfnamefont {M.~E.}\ \bibnamefont
  {Palumbo}}, \bibinfo {author} {\bibfnamefont {G.~A.}\ \bibnamefont
  {Baratta}}, \bibinfo {author} {\bibfnamefont {G.}~\bibnamefont {Leto}}, \
  and\ \bibinfo {author} {\bibfnamefont {G.}~\bibnamefont {Strazzulla}},\
  }\bibfield  {title} {\enquote {\bibinfo {title} {{H bonds in astrophysical
  ices}},}\ }\href {\doibase http://dx.doi.org/10.1016/j.molstruc.2009.12.017}
  {\bibfield  {journal} {\bibinfo  {journal} {Journal of Molecular Structure}\
  }\textbf {\bibinfo {volume} {972}},\ \bibinfo {pages} {64--67} (\bibinfo
  {year} {2010})}\BibitemShut {NoStop}%
\bibitem [{\citenamefont {{Mu{\~n}oz Caro}}\ \emph {et~al.}(2010)\citenamefont
  {{Mu{\~n}oz Caro}}, \citenamefont {{Jim{\'e}nez-Escobar}}, \citenamefont
  {{Mart{\'\i}n-Gago}}, \citenamefont {{Rogero}}, \citenamefont {{Atienza}},
  \citenamefont {{Puertas}}, \citenamefont {{Sobrado}},\ and\ \citenamefont
  {{Torres-Redondo}}}]{munozcaro10}%
  \BibitemOpen
  \bibfield  {author} {\bibinfo {author} {\bibfnamefont {G.~M.}\ \bibnamefont
  {{Mu{\~n}oz Caro}}}, \bibinfo {author} {\bibfnamefont {A.}~\bibnamefont
  {{Jim{\'e}nez-Escobar}}}, \bibinfo {author} {\bibfnamefont {J.~{\'A}.}\
  \bibnamefont {{Mart{\'\i}n-Gago}}}, \bibinfo {author} {\bibfnamefont
  {C.}~\bibnamefont {{Rogero}}}, \bibinfo {author} {\bibfnamefont
  {C.}~\bibnamefont {{Atienza}}}, \bibinfo {author} {\bibfnamefont
  {S.}~\bibnamefont {{Puertas}}}, \bibinfo {author} {\bibfnamefont {J.~M.}\
  \bibnamefont {{Sobrado}}}, \ and\ \bibinfo {author} {\bibfnamefont
  {J.}~\bibnamefont {{Torres-Redondo}}},\ }\bibfield  {title} {\enquote
  {\bibinfo {title} {{New results on thermal and photodesorption of CO ice
  using the novel InterStellar Astrochemistry Chamber (ISAC)}},}\ }\href
  {\doibase 10.1051/0004-6361/200912462} {\bibfield  {journal} {\bibinfo
  {journal} {Astronomy and Astrophysics}\ }\textbf {\bibinfo {volume} {522}},\
  \bibinfo {eid} {A108} (\bibinfo {year} {2010})}\BibitemShut {NoStop}%
\bibitem [{\citenamefont {Linnartz}, \citenamefont {Ioppolo},\ and\
  \citenamefont {Fedoseev}(2015)}]{linnartz15}%
  \BibitemOpen
  \bibfield  {author} {\bibinfo {author} {\bibfnamefont {H.}~\bibnamefont
  {Linnartz}}, \bibinfo {author} {\bibfnamefont {S.}~\bibnamefont {Ioppolo}}, \
  and\ \bibinfo {author} {\bibfnamefont {G.}~\bibnamefont {Fedoseev}},\
  }\bibfield  {title} {\enquote {\bibinfo {title} {Atom addition reactions in
  interstellar ice analogues},}\ }\href {\doibase
  10.1080/0144235X.2015.1046679} {\bibfield  {journal} {\bibinfo  {journal}
  {International Reviews in Physical Chemistry}\ }\textbf {\bibinfo {volume}
  {34}},\ \bibinfo {pages} {205--237} (\bibinfo {year} {2015})}\BibitemShut
  {NoStop}%
\bibitem [{\citenamefont {Fulvio}\ \emph {et~al.}(2017)\citenamefont {Fulvio},
  \citenamefont {Sandor}, \citenamefont {Jager}, \citenamefont {Akos},\ and\
  \citenamefont {Henning}}]{fulvio17}%
  \BibitemOpen
  \bibfield  {author} {\bibinfo {author} {\bibfnamefont {D.}~\bibnamefont
  {Fulvio}}, \bibinfo {author} {\bibfnamefont {G.}~\bibnamefont {Sandor}},
  \bibinfo {author} {\bibfnamefont {C.}~\bibnamefont {Jager}}, \bibinfo
  {author} {\bibfnamefont {K.}~\bibnamefont {Akos}}, \ and\ \bibinfo {author}
  {\bibfnamefont {T.}~\bibnamefont {Henning}},\ }\bibfield  {title} {\enquote
  {\bibinfo {title} {{Laboratory Experiments on the Low-temperature Formation
  of Carbonaceous Grains in the ISM}},}\ }\href {\doibase
  10.3847/1538-4365/aa9224} {\bibfield  {journal} {\bibinfo  {journal}
  {Astrophysical Journal Supplement Series}\ }\textbf {\bibinfo {volume}
  {233}},\ \bibinfo {pages} {11} (\bibinfo {year} {2017})}\BibitemShut
  {NoStop}%
\bibitem [{\citenamefont {Hudson}, \citenamefont {Loeffler},\ and\
  \citenamefont {Yocum}(2017)}]{hudson17}%
  \BibitemOpen
  \bibfield  {author} {\bibinfo {author} {\bibfnamefont {R.~L.}\ \bibnamefont
  {Hudson}}, \bibinfo {author} {\bibfnamefont {M.~J.}\ \bibnamefont
  {Loeffler}}, \ and\ \bibinfo {author} {\bibfnamefont {K.~M.}\ \bibnamefont
  {Yocum}},\ }\bibfield  {title} {\enquote {\bibinfo {title} {Laboratory
  investigations into the spectra and origin of propylene oxide: A chiral
  interstellar molecule},}\ }\href {\doibase 10.3847/1538-4357/835/2/225}
  {\bibfield  {journal} {\bibinfo  {journal} {The Astrophysical Journal}\
  }\textbf {\bibinfo {volume} {835}},\ \bibinfo {pages} {225} (\bibinfo {year}
  {2017})}\BibitemShut {NoStop}%
\bibitem [{\citenamefont {Potapov}, \citenamefont {Jäger},\ and\ \citenamefont
  {Henning}(2019)}]{potapov19a}%
  \BibitemOpen
  \bibfield  {author} {\bibinfo {author} {\bibfnamefont {A.}~\bibnamefont
  {Potapov}}, \bibinfo {author} {\bibfnamefont {C.}~\bibnamefont {Jäger}}, \
  and\ \bibinfo {author} {\bibfnamefont {T.}~\bibnamefont {Henning}},\
  }\bibfield  {title} {\enquote {\bibinfo {title} {Photodesorption of water ice
  from dust grains and thermal desorption of cometary ices studied by the
  {INSIDE} experiment},}\ }\href {\doibase 10.3847/1538-4357/ab25e7} {\bibfield
   {journal} {\bibinfo  {journal} {The Astrophysical Journal}\ }\textbf
  {\bibinfo {volume} {880}},\ \bibinfo {pages} {12} (\bibinfo {year}
  {2019})}\BibitemShut {NoStop}%
\bibitem [{\citenamefont {Oba}\ \emph {et~al.}(2019)\citenamefont {Oba},
  \citenamefont {Takano}, \citenamefont {Naraoka}, \citenamefont {Watanabe},\
  and\ \citenamefont {Kouchi}}]{oba19}%
  \BibitemOpen
  \bibfield  {author} {\bibinfo {author} {\bibfnamefont {Y.}~\bibnamefont
  {Oba}}, \bibinfo {author} {\bibfnamefont {Y.}~\bibnamefont {Takano}},
  \bibinfo {author} {\bibfnamefont {H.}~\bibnamefont {Naraoka}}, \bibinfo
  {author} {\bibfnamefont {N.}~\bibnamefont {Watanabe}}, \ and\ \bibinfo
  {author} {\bibfnamefont {A.}~\bibnamefont {Kouchi}},\ }\bibfield  {title}
  {\enquote {\bibinfo {title} {{Nucleobase synthesis in interstellar ices}},}\
  }\href {\doibase 10.1038/s41467-019-12404-1} {\bibfield  {journal} {\bibinfo
  {journal} {Nature Communications}\ }\textbf {\bibinfo {volume} {10}},\
  \bibinfo {pages} {4413} (\bibinfo {year} {2019})}\BibitemShut {NoStop}%
\bibitem [{\citenamefont {Boogert}, \citenamefont {Gerakines},\ and\
  \citenamefont {Whittet}(2015)}]{boogert15}%
  \BibitemOpen
  \bibfield  {author} {\bibinfo {author} {\bibfnamefont {A.~C.~A.}\
  \bibnamefont {Boogert}}, \bibinfo {author} {\bibfnamefont {P.~A.}\
  \bibnamefont {Gerakines}}, \ and\ \bibinfo {author} {\bibfnamefont
  {D.~C.~B.}\ \bibnamefont {Whittet}},\ }\bibfield  {title} {\enquote {\bibinfo
  {title} {{Observations of the Icy Universe}},}\ }\href {\doibase
  10.1146/annurev-astro-082214-122348} {\bibfield  {journal} {\bibinfo
  {journal} {Annual Review of Astronomy and Astrophysics, Vol 53}\ }\textbf
  {\bibinfo {volume} {53}},\ \bibinfo {pages} {541--581} (\bibinfo {year}
  {2015})}\BibitemShut {NoStop}%
\bibitem [{\citenamefont {{Pascoli}}\ and\ \citenamefont
  {{Polleux}}(2000)}]{pascoli00}%
  \BibitemOpen
  \bibfield  {author} {\bibinfo {author} {\bibfnamefont {G.}~\bibnamefont
  {{Pascoli}}}\ and\ \bibinfo {author} {\bibfnamefont {A.}~\bibnamefont
  {{Polleux}}},\ }\bibfield  {title} {\enquote {\bibinfo {title} {Condensation
  and growth of hydrogenated carbon clusters in carbon-rich stars},}\
  }\href@noop {} {\bibfield  {journal} {\bibinfo  {journal} {Astronomy \&
  Astrophysics}\ }\textbf {\bibinfo {volume} {359}},\ \bibinfo {pages}
  {799--810} (\bibinfo {year} {2000})}\BibitemShut {NoStop}%
\bibitem [{\citenamefont {Vidali}(2013)}]{vidali13}%
  \BibitemOpen
  \bibfield  {author} {\bibinfo {author} {\bibfnamefont {G.}~\bibnamefont
  {Vidali}},\ }\bibfield  {title} {\enquote {\bibinfo {title} {H2 formation on
  interstellar grains},}\ }\href {\doibase 10.1021/cr400156b} {\bibfield
  {journal} {\bibinfo  {journal} {Chemical Reviews}\ }\textbf {\bibinfo
  {volume} {113}},\ \bibinfo {pages} {8762--8782} (\bibinfo {year}
  {2013})}\BibitemShut {NoStop}%
\bibitem [{\citenamefont {Williams}\ and\ \citenamefont
  {Cecchi-Pestellini}(2016)}]{williams16}%
  \BibitemOpen
  \bibfield  {author} {\bibinfo {author} {\bibfnamefont {D.~A.}\ \bibnamefont
  {Williams}}\ and\ \bibinfo {author} {\bibfnamefont {C.}~\bibnamefont
  {Cecchi-Pestellini}},\ }\bibfield  {title} {\enquote {\bibinfo {title}
  {Chapter 8 catalysis on the surfaces of bare dust grains},}\ }in\ \href
  {\doibase 10.1039/9781782623694-00157} {\emph {\bibinfo {booktitle} {The
  Chemistry of Cosmic Dust}}}\ (\bibinfo  {publisher} {The Royal Society of
  Chemistry},\ \bibinfo {year} {2016})\ pp.\ \bibinfo {pages}
  {157--196}\BibitemShut {NoStop}%
\bibitem [{\citenamefont {Wakelam}\ \emph {et~al.}(2017)\citenamefont
  {Wakelam}, \citenamefont {Bron}, \citenamefont {Cazaux}, \citenamefont
  {Dulieu}, \citenamefont {Gry}, \citenamefont {Guillard}, \citenamefont
  {Habart}, \citenamefont {Hornekær}, \citenamefont {Morisset}, \citenamefont
  {Nyman}, \citenamefont {Pirronello}, \citenamefont {Price}, \citenamefont
  {Valdivia}, \citenamefont {Vidali},\ and\ \citenamefont
  {Watanabe}}]{wakelam17}%
  \BibitemOpen
  \bibfield  {author} {\bibinfo {author} {\bibfnamefont {V.}~\bibnamefont
  {Wakelam}}, \bibinfo {author} {\bibfnamefont {E.}~\bibnamefont {Bron}},
  \bibinfo {author} {\bibfnamefont {S.}~\bibnamefont {Cazaux}}, \bibinfo
  {author} {\bibfnamefont {F.}~\bibnamefont {Dulieu}}, \bibinfo {author}
  {\bibfnamefont {C.}~\bibnamefont {Gry}}, \bibinfo {author} {\bibfnamefont
  {P.}~\bibnamefont {Guillard}}, \bibinfo {author} {\bibfnamefont
  {E.}~\bibnamefont {Habart}}, \bibinfo {author} {\bibfnamefont
  {L.}~\bibnamefont {Hornekær}}, \bibinfo {author} {\bibfnamefont
  {S.}~\bibnamefont {Morisset}}, \bibinfo {author} {\bibfnamefont
  {G.}~\bibnamefont {Nyman}}, \bibinfo {author} {\bibfnamefont
  {V.}~\bibnamefont {Pirronello}}, \bibinfo {author} {\bibfnamefont {S.~D.}\
  \bibnamefont {Price}}, \bibinfo {author} {\bibfnamefont {V.}~\bibnamefont
  {Valdivia}}, \bibinfo {author} {\bibfnamefont {G.}~\bibnamefont {Vidali}}, \
  and\ \bibinfo {author} {\bibfnamefont {N.}~\bibnamefont {Watanabe}},\
  }\bibfield  {title} {\enquote {\bibinfo {title} {H$_2$ formation on
  interstellar dust grains: The viewpoints of theory, experiments, models and
  observations},}\ }\href {\doibase
  https://doi.org/10.1016/j.molap.2017.11.001} {\bibfield  {journal} {\bibinfo
  {journal} {Molecular Astrophysics}\ }\textbf {\bibinfo {volume} {9}},\
  \bibinfo {pages} {1 -- 36} (\bibinfo {year} {2017})}\BibitemShut {NoStop}%
\bibitem [{\citenamefont {Allamandola}\ \emph {et~al.}(1999)\citenamefont
  {Allamandola}, \citenamefont {Bernstein}, \citenamefont {Sandford},\ and\
  \citenamefont {Walker}}]{allamandola99}%
  \BibitemOpen
  \bibfield  {author} {\bibinfo {author} {\bibfnamefont {L.~J.}\ \bibnamefont
  {Allamandola}}, \bibinfo {author} {\bibfnamefont {M.~P.}\ \bibnamefont
  {Bernstein}}, \bibinfo {author} {\bibfnamefont {S.~A.}\ \bibnamefont
  {Sandford}}, \ and\ \bibinfo {author} {\bibfnamefont {R.~L.}\ \bibnamefont
  {Walker}},\ }\bibfield  {title} {\enquote {\bibinfo {title} {{Evolution of
  Interstellar Ices}},}\ }\href {\doibase 10.1023/a:1005210417396} {\bibfield
  {journal} {\bibinfo  {journal} {Space Science Reviews}\ }\textbf {\bibinfo
  {volume} {90}},\ \bibinfo {pages} {219--232} (\bibinfo {year}
  {1999})}\BibitemShut {NoStop}%
\bibitem [{\citenamefont {Strazzulla}, \citenamefont {Baratta},\ and\
  \citenamefont {Palumbo}(2001)}]{strazzulla01}%
  \BibitemOpen
  \bibfield  {author} {\bibinfo {author} {\bibfnamefont {G.}~\bibnamefont
  {Strazzulla}}, \bibinfo {author} {\bibfnamefont {G.}~\bibnamefont {Baratta}},
  \ and\ \bibinfo {author} {\bibfnamefont {M.}~\bibnamefont {Palumbo}},\
  }\bibfield  {title} {\enquote {\bibinfo {title} {Vibrational spectroscopy of
  ion-irradiated ices},}\ }\href {\doibase 10.1016/S1386-1425(00)00447-9}
  {\bibfield  {journal} {\bibinfo  {journal} {Spectrochimica Acta Part A:
  Molecular and Biomolecular Spectroscopy}\ }\textbf {\bibinfo {volume} {57}},\
  \bibinfo {pages} {825 -- 842} (\bibinfo {year} {2001})}\BibitemShut {NoStop}%
\bibitem [{\citenamefont {Johnson}(1998)}]{johnson98}%
  \BibitemOpen
  \bibfield  {author} {\bibinfo {author} {\bibfnamefont {R.~E.}\ \bibnamefont
  {Johnson}},\ }\enquote {\bibinfo {title} {Sputtering and desorption from icy
  surfaces},}\ in\ \href {\doibase 10.1007/978-94-011-5252-5_13} {\emph
  {\bibinfo {booktitle} {Solar System Ices}}},\ \bibinfo {editor} {edited by\
  \bibinfo {editor} {\bibfnamefont {B.}~\bibnamefont {Schmitt}}, \bibinfo
  {editor} {\bibfnamefont {C.}~\bibnamefont {De~Bergh}}, \ and\ \bibinfo
  {editor} {\bibfnamefont {M.}~\bibnamefont {Festou}}}\ (\bibinfo  {publisher}
  {Springer Netherlands},\ \bibinfo {address} {Dordrecht},\ \bibinfo {year}
  {1998})\ pp.\ \bibinfo {pages} {303--334}\BibitemShut {NoStop}%
\bibitem [{\citenamefont {Öberg}(2016)}]{oberg16}%
  \BibitemOpen
  \bibfield  {author} {\bibinfo {author} {\bibfnamefont {K.~I.}\ \bibnamefont
  {Öberg}},\ }\bibfield  {title} {\enquote {\bibinfo {title} {Photochemistry
  and astrochemistry: Photochemical pathways to interstellar complex organic
  molecules},}\ }\href {\doibase 10.1021/acs.chemrev.5b00694} {\bibfield
  {journal} {\bibinfo  {journal} {Chemical Reviews}\ }\textbf {\bibinfo
  {volume} {116}},\ \bibinfo {pages} {9631--9663} (\bibinfo {year}
  {2016})}\BibitemShut {NoStop}%
\bibitem [{\citenamefont {{Mu{\~{n}}oz Caro}}\ \emph
  {et~al.}(2002)\citenamefont {{Mu{\~{n}}oz Caro}}, \citenamefont
  {Meierhenrich}, \citenamefont {Schutte}, \citenamefont {Barbier},
  \citenamefont {{Arcones Segovia}}, \citenamefont {Rosenbauer}, \citenamefont
  {Thiemann}, \citenamefont {Brack},\ and\ \citenamefont
  {Greenberg}}]{munozcaro02}%
  \BibitemOpen
  \bibfield  {author} {\bibinfo {author} {\bibfnamefont {G.~M.}\ \bibnamefont
  {{Mu{\~{n}}oz Caro}}}, \bibinfo {author} {\bibfnamefont {U.~J.}\ \bibnamefont
  {Meierhenrich}}, \bibinfo {author} {\bibfnamefont {W.~A.}\ \bibnamefont
  {Schutte}}, \bibinfo {author} {\bibfnamefont {B.}~\bibnamefont {Barbier}},
  \bibinfo {author} {\bibfnamefont {A.}~\bibnamefont {{Arcones Segovia}}},
  \bibinfo {author} {\bibfnamefont {H.}~\bibnamefont {Rosenbauer}}, \bibinfo
  {author} {\bibfnamefont {W.~H.-P.}\ \bibnamefont {Thiemann}}, \bibinfo
  {author} {\bibfnamefont {A.}~\bibnamefont {Brack}}, \ and\ \bibinfo {author}
  {\bibfnamefont {J.~M.}\ \bibnamefont {Greenberg}},\ }\bibfield  {title}
  {\enquote {\bibinfo {title} {{Amino acids from ultraviolet irradiation of
  interstellar ice analogues}},}\ }\href {\doibase 10.1038/416403a} {\bibfield
  {journal} {\bibinfo  {journal} {Nature}\ }\textbf {\bibinfo {volume} {416}},\
  \bibinfo {pages} {403--406} (\bibinfo {year} {2002})}\BibitemShut {NoStop}%
\bibitem [{\citenamefont {Meinert}\ \emph {et~al.}(2016)\citenamefont
  {Meinert}, \citenamefont {Myrgorodska}, \citenamefont {de~Marcellus},
  \citenamefont {Buhse}, \citenamefont {Nahon}, \citenamefont {Hoffmann},
  \citenamefont {d{\textquoteright}Hendecourt},\ and\ \citenamefont
  {Meierhenrich}}]{meinert16}%
  \BibitemOpen
  \bibfield  {author} {\bibinfo {author} {\bibfnamefont {C.}~\bibnamefont
  {Meinert}}, \bibinfo {author} {\bibfnamefont {I.}~\bibnamefont
  {Myrgorodska}}, \bibinfo {author} {\bibfnamefont {P.}~\bibnamefont
  {de~Marcellus}}, \bibinfo {author} {\bibfnamefont {T.}~\bibnamefont {Buhse}},
  \bibinfo {author} {\bibfnamefont {L.}~\bibnamefont {Nahon}}, \bibinfo
  {author} {\bibfnamefont {S.~V.}\ \bibnamefont {Hoffmann}}, \bibinfo {author}
  {\bibfnamefont {L.~L.~S.}\ \bibnamefont {d{\textquoteright}Hendecourt}}, \
  and\ \bibinfo {author} {\bibfnamefont {U.~J.}\ \bibnamefont {Meierhenrich}},\
  }\bibfield  {title} {\enquote {\bibinfo {title} {Ribose and related sugars
  from ultraviolet irradiation of interstellar ice analogs},}\ }\href {\doibase
  10.1126/science.aad8137} {\bibfield  {journal} {\bibinfo  {journal}
  {Science}\ }\textbf {\bibinfo {volume} {352}},\ \bibinfo {pages} {208--212}
  (\bibinfo {year} {2016})}\BibitemShut {NoStop}%
\bibitem [{\citenamefont {Tarczay}\ \emph {et~al.}(2016)\citenamefont
  {Tarczay}, \citenamefont {Förstel}, \citenamefont {Maksyutenko},\ and\
  \citenamefont {Kaiser}}]{tarczay16}%
  \BibitemOpen
  \bibfield  {author} {\bibinfo {author} {\bibfnamefont {G.}~\bibnamefont
  {Tarczay}}, \bibinfo {author} {\bibfnamefont {M.}~\bibnamefont {Förstel}},
  \bibinfo {author} {\bibfnamefont {P.}~\bibnamefont {Maksyutenko}}, \ and\
  \bibinfo {author} {\bibfnamefont {R.~I.}\ \bibnamefont {Kaiser}},\ }\bibfield
   {title} {\enquote {\bibinfo {title} {Formation of higher silanes in
  low-temperature silane {(SiH4)} ices},}\ }\href {\doibase
  10.1021/acs.inorgchem.6b01327} {\bibfield  {journal} {\bibinfo  {journal}
  {Inorganic Chemistry}\ }\textbf {\bibinfo {volume} {55}},\ \bibinfo {pages}
  {8776--8785} (\bibinfo {year} {2016})}\BibitemShut {NoStop}%
\bibitem [{\citenamefont {Esmaili}\ \emph {et~al.}(2018)\citenamefont
  {Esmaili}, \citenamefont {Bass}, \citenamefont {Cloutier}, \citenamefont
  {Sanche},\ and\ \citenamefont {Huels}}]{esmaili18}%
  \BibitemOpen
  \bibfield  {author} {\bibinfo {author} {\bibfnamefont {S.}~\bibnamefont
  {Esmaili}}, \bibinfo {author} {\bibfnamefont {A.~D.}\ \bibnamefont {Bass}},
  \bibinfo {author} {\bibfnamefont {P.}~\bibnamefont {Cloutier}}, \bibinfo
  {author} {\bibfnamefont {L.}~\bibnamefont {Sanche}}, \ and\ \bibinfo {author}
  {\bibfnamefont {M.~A.}\ \bibnamefont {Huels}},\ }\bibfield  {title} {\enquote
  {\bibinfo {title} {Glycine formation in co2:ch4:nh3 ices induced by 0-70 ev
  electrons},}\ }\href {\doibase 10.1063/1.5021596} {\bibfield  {journal}
  {\bibinfo  {journal} {The Journal of Chemical Physics}\ }\textbf {\bibinfo
  {volume} {148}},\ \bibinfo {pages} {164702} (\bibinfo {year}
  {2018})}\BibitemShut {NoStop}%
\bibitem [{\citenamefont {Merino}\ \emph {et~al.}(2014)\citenamefont {Merino},
  \citenamefont {Švec}, \citenamefont {Martinez}, \citenamefont {Jelinek},
  \citenamefont {Lacovig}, \citenamefont {Dalmiglio}, \citenamefont {Lizzit},
  \citenamefont {Soukiassian}, \citenamefont {Cernicharo},\ and\ \citenamefont
  {Martin-Gago}}]{merino14}%
  \BibitemOpen
  \bibfield  {author} {\bibinfo {author} {\bibfnamefont {P.}~\bibnamefont
  {Merino}}, \bibinfo {author} {\bibfnamefont {M.}~\bibnamefont {Švec}},
  \bibinfo {author} {\bibfnamefont {J.~I.}\ \bibnamefont {Martinez}}, \bibinfo
  {author} {\bibfnamefont {P.}~\bibnamefont {Jelinek}}, \bibinfo {author}
  {\bibfnamefont {P.}~\bibnamefont {Lacovig}}, \bibinfo {author} {\bibfnamefont
  {M.}~\bibnamefont {Dalmiglio}}, \bibinfo {author} {\bibfnamefont
  {S.}~\bibnamefont {Lizzit}}, \bibinfo {author} {\bibfnamefont
  {P.}~\bibnamefont {Soukiassian}}, \bibinfo {author} {\bibfnamefont
  {J.}~\bibnamefont {Cernicharo}}, \ and\ \bibinfo {author} {\bibfnamefont
  {J.~A.}\ \bibnamefont {Martin-Gago}},\ }\bibfield  {title} {\enquote
  {\bibinfo {title} {Graphene etching on {SiC} grains as a path to interstellar
  polycyclic aromatic hydrocarbons formation},}\ }\href
  {https://doi.org/10.1038/ncomms4054} {\bibfield  {journal} {\bibinfo
  {journal} {Nature Communications}\ }\textbf {\bibinfo {volume} {5}},\
  \bibinfo {pages} {3054} (\bibinfo {year} {2014})}\BibitemShut {NoStop}%
\bibitem [{\citenamefont {Fraser}\ and\ \citenamefont {{van
  Dishoeck}}(2004)}]{fraser04}%
  \BibitemOpen
  \bibfield  {author} {\bibinfo {author} {\bibfnamefont {H.}~\bibnamefont
  {Fraser}}\ and\ \bibinfo {author} {\bibfnamefont {E.}~\bibnamefont {{van
  Dishoeck}}},\ }\bibfield  {title} {\enquote {\bibinfo {title} {{SURFRESIDE}:
  a novel experiment to study surface chemistry under interstellar and
  protostellar conditions},}\ }\href {\doibase
  https://doi.org/10.1016/j.asr.2003.04.003} {\bibfield  {journal} {\bibinfo
  {journal} {Advances in Space Research}\ }\textbf {\bibinfo {volume} {33}},\
  \bibinfo {pages} {14 -- 22} (\bibinfo {year} {2004})}\BibitemShut {NoStop}%
\bibitem [{\citenamefont {Ioppolo}\ \emph {et~al.}(2013)\citenamefont
  {Ioppolo}, \citenamefont {Fedoseev}, \citenamefont {Lamberts}, \citenamefont
  {Romanzin},\ and\ \citenamefont {Linnartz}}]{ioppolo13}%
  \BibitemOpen
  \bibfield  {author} {\bibinfo {author} {\bibfnamefont {S.}~\bibnamefont
  {Ioppolo}}, \bibinfo {author} {\bibfnamefont {G.}~\bibnamefont {Fedoseev}},
  \bibinfo {author} {\bibfnamefont {T.}~\bibnamefont {Lamberts}}, \bibinfo
  {author} {\bibfnamefont {C.}~\bibnamefont {Romanzin}}, \ and\ \bibinfo
  {author} {\bibfnamefont {H.}~\bibnamefont {Linnartz}},\ }\bibfield  {title}
  {\enquote {\bibinfo {title} {{SURFRESIDE2}: An ultrahigh vacuum system for
  the investigation of surface reaction routes of interstellar interest},}\
  }\href {\doibase 10.1063/1.4816135} {\bibfield  {journal} {\bibinfo
  {journal} {Review of Scientific Instruments}\ }\textbf {\bibinfo {volume}
  {84}},\ \bibinfo {pages} {073112} (\bibinfo {year} {2013})}\BibitemShut
  {NoStop}%
\bibitem [{\citenamefont {Potapov}\ \emph {et~al.}(2019)\citenamefont
  {Potapov}, \citenamefont {Theul{\'{e}}}, \citenamefont {Jäger},\ and\
  \citenamefont {Henning}}]{potapov19b}%
  \BibitemOpen
  \bibfield  {author} {\bibinfo {author} {\bibfnamefont {A.}~\bibnamefont
  {Potapov}}, \bibinfo {author} {\bibfnamefont {P.}~\bibnamefont
  {Theul{\'{e}}}}, \bibinfo {author} {\bibfnamefont {C.}~\bibnamefont
  {Jäger}}, \ and\ \bibinfo {author} {\bibfnamefont {T.}~\bibnamefont
  {Henning}},\ }\bibfield  {title} {\enquote {\bibinfo {title} {Evidence of
  surface catalytic effect on cosmic dust grain analogs: The ammonia and carbon
  dioxide surface reaction},}\ }\href {\doibase 10.3847/2041-8213/ab2538}
  {\bibfield  {journal} {\bibinfo  {journal} {The Astrophysical Journal}\
  }\textbf {\bibinfo {volume} {878}},\ \bibinfo {pages} {L20} (\bibinfo {year}
  {2019})}\BibitemShut {NoStop}%
\bibitem [{\citenamefont {Potapov}, \citenamefont {J\"ager},\ and\
  \citenamefont {Henning}(2020)}]{potapov20}%
  \BibitemOpen
  \bibfield  {author} {\bibinfo {author} {\bibfnamefont {A.}~\bibnamefont
  {Potapov}}, \bibinfo {author} {\bibfnamefont {C.}~\bibnamefont {J\"ager}}, \
  and\ \bibinfo {author} {\bibfnamefont {T.}~\bibnamefont {Henning}},\
  }\bibfield  {title} {\enquote {\bibinfo {title} {Ice coverage of dust grains
  in cold astrophysical environments},}\ }\href {\doibase
  10.1103/PhysRevLett.124.221103} {\bibfield  {journal} {\bibinfo  {journal}
  {Phys. Rev. Lett.}\ }\textbf {\bibinfo {volume} {124}},\ \bibinfo {pages}
  {221103} (\bibinfo {year} {2020})}\BibitemShut {NoStop}%
\bibitem [{\citenamefont {Mart\'i{}nez}\ \emph {et~al.}(2018)\citenamefont
  {Mart\'i{}nez}, \citenamefont {Lauwaet}, \citenamefont {Santoro},
  \citenamefont {Sobrado}, \citenamefont {Pel\'aez}, \citenamefont {Herrero},
  \citenamefont {Tanarro}, \citenamefont {Ellis}, \citenamefont {Cernicharo},
  \citenamefont {Joblin}, \citenamefont {Huttel},\ and\ \citenamefont
  {Mart\'i{}n-Gago}}]{martinez18}%
  \BibitemOpen
  \bibfield  {author} {\bibinfo {author} {\bibfnamefont {L.}~\bibnamefont
  {Mart\'i{}nez}}, \bibinfo {author} {\bibfnamefont {K.}~\bibnamefont
  {Lauwaet}}, \bibinfo {author} {\bibfnamefont {G.}~\bibnamefont {Santoro}},
  \bibinfo {author} {\bibfnamefont {J.~M.}\ \bibnamefont {Sobrado}}, \bibinfo
  {author} {\bibfnamefont {R.~J.}\ \bibnamefont {Pel\'aez}}, \bibinfo {author}
  {\bibfnamefont {V.~J.}\ \bibnamefont {Herrero}}, \bibinfo {author}
  {\bibfnamefont {I.}~\bibnamefont {Tanarro}}, \bibinfo {author} {\bibfnamefont
  {G.~J.}\ \bibnamefont {Ellis}}, \bibinfo {author} {\bibfnamefont
  {J.}~\bibnamefont {Cernicharo}}, \bibinfo {author} {\bibfnamefont
  {C.}~\bibnamefont {Joblin}}, \bibinfo {author} {\bibfnamefont
  {Y.}~\bibnamefont {Huttel}}, \ and\ \bibinfo {author} {\bibfnamefont {J.~A.}\
  \bibnamefont {Mart\'i{}n-Gago}},\ }\bibfield  {title} {\enquote {\bibinfo
  {title} {Precisely controlled fabrication, manipulation and in-situ analysis
  of {Cu} based nanoparticles},}\ }\href
  {https://doi.org/10.1038/s41598-018-25472-y} {\bibfield  {journal} {\bibinfo
  {journal} {Scientific Reports}\ }\textbf {\bibinfo {volume} {8}},\ \bibinfo
  {pages} {7250} (\bibinfo {year} {2018})}\BibitemShut {NoStop}%
\bibitem [{\citenamefont {Chiar}\ \emph {et~al.}(2000)\citenamefont {Chiar},
  \citenamefont {Tielens}, \citenamefont {Whittet}, \citenamefont {Schutte},
  \citenamefont {Boogert}, \citenamefont {Lutz}, \citenamefont {van Dishoeck},\
  and\ \citenamefont {Bernstein}}]{chiar00}%
  \BibitemOpen
  \bibfield  {author} {\bibinfo {author} {\bibfnamefont {J.~E.}\ \bibnamefont
  {Chiar}}, \bibinfo {author} {\bibfnamefont {A.~G. G.~M.}\ \bibnamefont
  {Tielens}}, \bibinfo {author} {\bibfnamefont {D.~C.~B.}\ \bibnamefont
  {Whittet}}, \bibinfo {author} {\bibfnamefont {W.~A.}\ \bibnamefont
  {Schutte}}, \bibinfo {author} {\bibfnamefont {A.~C.~A.}\ \bibnamefont
  {Boogert}}, \bibinfo {author} {\bibfnamefont {D.}~\bibnamefont {Lutz}},
  \bibinfo {author} {\bibfnamefont {E.~F.}\ \bibnamefont {van Dishoeck}}, \
  and\ \bibinfo {author} {\bibfnamefont {M.~P.}\ \bibnamefont {Bernstein}},\
  }\bibfield  {title} {\enquote {\bibinfo {title} {The composition and
  distribution of dust along the line of sight toward the galactic center},}\
  }\href {\doibase 10.1086/309047} {\bibfield  {journal} {\bibinfo  {journal}
  {The Astrophysical Journal}\ }\textbf {\bibinfo {volume} {537}},\ \bibinfo
  {pages} {749--762} (\bibinfo {year} {2000})}\BibitemShut {NoStop}%
\bibitem [{\citenamefont {Pendleton}\ and\ \citenamefont
  {Allamandola}(2002)}]{pendleton02}%
  \BibitemOpen
  \bibfield  {author} {\bibinfo {author} {\bibfnamefont {Y.~J.}\ \bibnamefont
  {Pendleton}}\ and\ \bibinfo {author} {\bibfnamefont {L.~J.}\ \bibnamefont
  {Allamandola}},\ }\bibfield  {title} {\enquote {\bibinfo {title} {The organic
  refractory material in the diffuse interstellar medium: Mid-infrared
  spectroscopic constraints},}\ }\href {\doibase 10.1086/322999} {\bibfield
  {journal} {\bibinfo  {journal} {The Astrophysical Journal Supplement Series}\
  }\textbf {\bibinfo {volume} {138}},\ \bibinfo {pages} {75--98} (\bibinfo
  {year} {2002})}\BibitemShut {NoStop}%
\bibitem [{\citenamefont {Günay}\ \emph {et~al.}(2020)\citenamefont {Günay},
  \citenamefont {Burton}, \citenamefont {Afşar},\ and\ \citenamefont
  {Schmidt}}]{gunay20}%
  \BibitemOpen
  \bibfield  {author} {\bibinfo {author} {\bibfnamefont {B.}~\bibnamefont
  {Günay}}, \bibinfo {author} {\bibfnamefont {M.~G.}\ \bibnamefont {Burton}},
  \bibinfo {author} {\bibfnamefont {M.}~\bibnamefont {Afşar}}, \ and\ \bibinfo
  {author} {\bibfnamefont {T.~W.}\ \bibnamefont {Schmidt}},\ }\bibfield
  {title} {\enquote {\bibinfo {title} {{A method for mapping the aliphatic
  hydrocarbon content of interstellar dust towards the Galactic Centre}},}\
  }\href {\doibase 10.1093/mnras/staa288} {\bibfield  {journal} {\bibinfo
  {journal} {Monthly Notices of the Royal Astronomical Society}\ }\textbf
  {\bibinfo {volume} {493}},\ \bibinfo {pages} {1109--1119} (\bibinfo {year}
  {2020})}\BibitemShut {NoStop}%
\bibitem [{\citenamefont {Keller}\ \emph {et~al.}(2006)\citenamefont {Keller},
  \citenamefont {Bajt}, \citenamefont {Baratta}, \citenamefont {Borg},
  \citenamefont {Bradley}, \citenamefont {Brownlee}, \citenamefont {Busemann},
  \citenamefont {Brucato}, \citenamefont {Burchell}, \citenamefont {Colangeli},
  \citenamefont {d{\textquoteright}Hendecourt}, \citenamefont {Djouadi},
  \citenamefont {Ferrini}, \citenamefont {Flynn}, \citenamefont {Franchi},
  \citenamefont {Fries}, \citenamefont {Grady}, \citenamefont {Graham},
  \citenamefont {Grossemy}, \citenamefont {Kearsley}, \citenamefont {Matrajt},
  \citenamefont {Nakamura-Messenger}, \citenamefont {Mennella}, \citenamefont
  {Nittler}, \citenamefont {Palumbo}, \citenamefont {Stadermann}, \citenamefont
  {Tsou}, \citenamefont {Rotundi}, \citenamefont {Sandford}, \citenamefont
  {Snead}, \citenamefont {Steele}, \citenamefont {Wooden},\ and\ \citenamefont
  {Zolensky}}]{keller06}%
  \BibitemOpen
  \bibfield  {author} {\bibinfo {author} {\bibfnamefont {L.~P.}\ \bibnamefont
  {Keller}}, \bibinfo {author} {\bibfnamefont {S.}~\bibnamefont {Bajt}},
  \bibinfo {author} {\bibfnamefont {G.~A.}\ \bibnamefont {Baratta}}, \bibinfo
  {author} {\bibfnamefont {J.}~\bibnamefont {Borg}}, \bibinfo {author}
  {\bibfnamefont {J.~P.}\ \bibnamefont {Bradley}}, \bibinfo {author}
  {\bibfnamefont {D.~E.}\ \bibnamefont {Brownlee}}, \bibinfo {author}
  {\bibfnamefont {H.}~\bibnamefont {Busemann}}, \bibinfo {author}
  {\bibfnamefont {J.~R.}\ \bibnamefont {Brucato}}, \bibinfo {author}
  {\bibfnamefont {M.}~\bibnamefont {Burchell}}, \bibinfo {author}
  {\bibfnamefont {L.}~\bibnamefont {Colangeli}}, \bibinfo {author}
  {\bibfnamefont {L.}~\bibnamefont {d{\textquoteright}Hendecourt}}, \bibinfo
  {author} {\bibfnamefont {Z.}~\bibnamefont {Djouadi}}, \bibinfo {author}
  {\bibfnamefont {G.}~\bibnamefont {Ferrini}}, \bibinfo {author} {\bibfnamefont
  {G.}~\bibnamefont {Flynn}}, \bibinfo {author} {\bibfnamefont {I.~A.}\
  \bibnamefont {Franchi}}, \bibinfo {author} {\bibfnamefont {M.}~\bibnamefont
  {Fries}}, \bibinfo {author} {\bibfnamefont {M.~M.}\ \bibnamefont {Grady}},
  \bibinfo {author} {\bibfnamefont {G.~A.}\ \bibnamefont {Graham}}, \bibinfo
  {author} {\bibfnamefont {F.}~\bibnamefont {Grossemy}}, \bibinfo {author}
  {\bibfnamefont {A.}~\bibnamefont {Kearsley}}, \bibinfo {author}
  {\bibfnamefont {G.}~\bibnamefont {Matrajt}}, \bibinfo {author} {\bibfnamefont
  {K.}~\bibnamefont {Nakamura-Messenger}}, \bibinfo {author} {\bibfnamefont
  {V.}~\bibnamefont {Mennella}}, \bibinfo {author} {\bibfnamefont
  {L.}~\bibnamefont {Nittler}}, \bibinfo {author} {\bibfnamefont {M.~E.}\
  \bibnamefont {Palumbo}}, \bibinfo {author} {\bibfnamefont {F.~J.}\
  \bibnamefont {Stadermann}}, \bibinfo {author} {\bibfnamefont
  {P.}~\bibnamefont {Tsou}}, \bibinfo {author} {\bibfnamefont {A.}~\bibnamefont
  {Rotundi}}, \bibinfo {author} {\bibfnamefont {S.~A.}\ \bibnamefont
  {Sandford}}, \bibinfo {author} {\bibfnamefont {C.}~\bibnamefont {Snead}},
  \bibinfo {author} {\bibfnamefont {A.}~\bibnamefont {Steele}}, \bibinfo
  {author} {\bibfnamefont {D.}~\bibnamefont {Wooden}}, \ and\ \bibinfo {author}
  {\bibfnamefont {M.}~\bibnamefont {Zolensky}},\ }\bibfield  {title} {\enquote
  {\bibinfo {title} {Infrared spectroscopy of comet 81p/wild 2 samples returned
  by stardust},}\ }\href {\doibase 10.1126/science.1135796} {\bibfield
  {journal} {\bibinfo  {journal} {Science}\ }\textbf {\bibinfo {volume}
  {314}},\ \bibinfo {pages} {1728--1731} (\bibinfo {year} {2006})}\BibitemShut
  {NoStop}%
\bibitem [{\citenamefont {Raponi}\ \emph {et~al.}(2020)\citenamefont {Raponi},
  \citenamefont {Ciarniello}, \citenamefont {Capaccioni}, \citenamefont
  {Mennella}, \citenamefont {Filacchione}, \citenamefont {Vinogradoff},
  \citenamefont {Poch}, \citenamefont {Beck}, \citenamefont {Quirico},
  \citenamefont {{De Sanctis}}, \citenamefont {Moroz}, \citenamefont {Kappel},
  \citenamefont {Erard}, \citenamefont {Bockel{\'{e}}e-Morvan}, \citenamefont
  {Longobardo}, \citenamefont {Tosi}, \citenamefont {Palomba}, \citenamefont
  {Combe}, \citenamefont {Rousseau}, \citenamefont {Arnold}, \citenamefont
  {Carlson}, \citenamefont {Pommerol}, \citenamefont {Pilorget}, \citenamefont
  {Fornasier}, \citenamefont {Bellucci}, \citenamefont {Barucci}, \citenamefont
  {Mancarella}, \citenamefont {Formisano}, \citenamefont {Rinaldi},
  \citenamefont {Istiqomah},\ and\ \citenamefont {Leyrat}}]{raponi20}%
  \BibitemOpen
  \bibfield  {author} {\bibinfo {author} {\bibfnamefont {A.}~\bibnamefont
  {Raponi}}, \bibinfo {author} {\bibfnamefont {M.}~\bibnamefont {Ciarniello}},
  \bibinfo {author} {\bibfnamefont {F.}~\bibnamefont {Capaccioni}}, \bibinfo
  {author} {\bibfnamefont {V.}~\bibnamefont {Mennella}}, \bibinfo {author}
  {\bibfnamefont {G.}~\bibnamefont {Filacchione}}, \bibinfo {author}
  {\bibfnamefont {V.}~\bibnamefont {Vinogradoff}}, \bibinfo {author}
  {\bibfnamefont {O.}~\bibnamefont {Poch}}, \bibinfo {author} {\bibfnamefont
  {P.}~\bibnamefont {Beck}}, \bibinfo {author} {\bibfnamefont {E.}~\bibnamefont
  {Quirico}}, \bibinfo {author} {\bibfnamefont {M.~C.}\ \bibnamefont {{De
  Sanctis}}}, \bibinfo {author} {\bibfnamefont {L.~V.}\ \bibnamefont {Moroz}},
  \bibinfo {author} {\bibfnamefont {D.}~\bibnamefont {Kappel}}, \bibinfo
  {author} {\bibfnamefont {S.}~\bibnamefont {Erard}}, \bibinfo {author}
  {\bibfnamefont {D.}~\bibnamefont {Bockel{\'{e}}e-Morvan}}, \bibinfo {author}
  {\bibfnamefont {A.}~\bibnamefont {Longobardo}}, \bibinfo {author}
  {\bibfnamefont {F.}~\bibnamefont {Tosi}}, \bibinfo {author} {\bibfnamefont
  {E.}~\bibnamefont {Palomba}}, \bibinfo {author} {\bibfnamefont {J.-P.}\
  \bibnamefont {Combe}}, \bibinfo {author} {\bibfnamefont {B.}~\bibnamefont
  {Rousseau}}, \bibinfo {author} {\bibfnamefont {G.}~\bibnamefont {Arnold}},
  \bibinfo {author} {\bibfnamefont {R.~W.}\ \bibnamefont {Carlson}}, \bibinfo
  {author} {\bibfnamefont {A.}~\bibnamefont {Pommerol}}, \bibinfo {author}
  {\bibfnamefont {C.}~\bibnamefont {Pilorget}}, \bibinfo {author}
  {\bibfnamefont {S.}~\bibnamefont {Fornasier}}, \bibinfo {author}
  {\bibfnamefont {G.}~\bibnamefont {Bellucci}}, \bibinfo {author}
  {\bibfnamefont {A.}~\bibnamefont {Barucci}}, \bibinfo {author} {\bibfnamefont
  {F.}~\bibnamefont {Mancarella}}, \bibinfo {author} {\bibfnamefont
  {M.}~\bibnamefont {Formisano}}, \bibinfo {author} {\bibfnamefont
  {G.}~\bibnamefont {Rinaldi}}, \bibinfo {author} {\bibfnamefont
  {I.}~\bibnamefont {Istiqomah}}, \ and\ \bibinfo {author} {\bibfnamefont
  {C.}~\bibnamefont {Leyrat}},\ }\bibfield  {title} {\enquote {\bibinfo {title}
  {{Infrared detection of aliphatic organics on a cometary nucleus}},}\ }\href
  {\doibase 10.1038/s41550-019-0992-8} {\bibfield  {journal} {\bibinfo
  {journal} {Nature Astronomy}\ }\textbf {\bibinfo {volume} {4}},\ \bibinfo
  {pages} {500--505} (\bibinfo {year} {2020})}\BibitemShut {NoStop}%
\bibitem [{\citenamefont {{Schuhmann}}\ \emph {et~al.}(2019)\citenamefont
  {{Schuhmann}}, \citenamefont {{Altwegg}}, \citenamefont {{Balsiger}},
  \citenamefont {{Berthelier}}, \citenamefont {{De Keyser}}, \citenamefont
  {{Fiethe}}, \citenamefont {{Fuselier}}, \citenamefont {{Gasc}}, \citenamefont
  {{Gombosi}}, \citenamefont {{H{\"a}nni}}, \citenamefont {{Rubin}},
  \citenamefont {{Tzou}},\ and\ \citenamefont {{Wampfler}}}]{schuhmann19}%
  \BibitemOpen
  \bibfield  {author} {\bibinfo {author} {\bibfnamefont {M.}~\bibnamefont
  {{Schuhmann}}}, \bibinfo {author} {\bibfnamefont {K.}~\bibnamefont
  {{Altwegg}}}, \bibinfo {author} {\bibfnamefont {H.}~\bibnamefont
  {{Balsiger}}}, \bibinfo {author} {\bibfnamefont {J.~J.}\ \bibnamefont
  {{Berthelier}}}, \bibinfo {author} {\bibfnamefont {J.}~\bibnamefont {{De
  Keyser}}}, \bibinfo {author} {\bibfnamefont {B.}~\bibnamefont {{Fiethe}}},
  \bibinfo {author} {\bibfnamefont {S.~A.}\ \bibnamefont {{Fuselier}}},
  \bibinfo {author} {\bibfnamefont {S.}~\bibnamefont {{Gasc}}}, \bibinfo
  {author} {\bibfnamefont {T.~I.}\ \bibnamefont {{Gombosi}}}, \bibinfo {author}
  {\bibfnamefont {N.}~\bibnamefont {{H{\"a}nni}}}, \bibinfo {author}
  {\bibfnamefont {M.}~\bibnamefont {{Rubin}}}, \bibinfo {author} {\bibfnamefont
  {C.~Y.}\ \bibnamefont {{Tzou}}}, \ and\ \bibinfo {author} {\bibfnamefont
  {S.~F.}\ \bibnamefont {{Wampfler}}},\ }\bibfield  {title} {\enquote {\bibinfo
  {title} {{Aliphatic and aromatic hydrocarbons in comet
  67P/Churyumov-Gerasimenko seen by ROSINA}},}\ }\href {\doibase
  10.1051/0004-6361/201834666} {\bibfield  {journal} {\bibinfo  {journal}
  {Astronomy and Astrophysics}\ }\textbf {\bibinfo {volume} {630}},\ \bibinfo
  {eid} {A31} (\bibinfo {year} {2019})}\BibitemShut {NoStop}%
\bibitem [{\citenamefont {Haberland}, \citenamefont {Karrais},\ and\
  \citenamefont {Mall}(1991)}]{haberland91}%
  \BibitemOpen
  \bibfield  {author} {\bibinfo {author} {\bibfnamefont {H.}~\bibnamefont
  {Haberland}}, \bibinfo {author} {\bibfnamefont {M.}~\bibnamefont {Karrais}},
  \ and\ \bibinfo {author} {\bibfnamefont {M.}~\bibnamefont {Mall}},\
  }\bibfield  {title} {\enquote {\bibinfo {title} {A new type of cluster and
  cluster ion source},}\ }\href {\doibase 10.1007/BF01544025} {\bibfield
  {journal} {\bibinfo  {journal} {Zeitschrift f{\"u}r Physik D Atoms, Molecules
  and Clusters}\ }\textbf {\bibinfo {volume} {20}},\ \bibinfo {pages}
  {413--415} (\bibinfo {year} {1991})}\BibitemShut {NoStop}%
\bibitem [{\citenamefont {Mart\'i{}nez}\ \emph {et~al.}(2012)\citenamefont
  {Mart\'i{}nez}, \citenamefont {D\'i{}az}, \citenamefont {Rom\'an},
  \citenamefont {Ruano}, \citenamefont {Llamosa~P.},\ and\ \citenamefont
  {Huttel}}]{martinez12}%
  \BibitemOpen
  \bibfield  {author} {\bibinfo {author} {\bibfnamefont {L.}~\bibnamefont
  {Mart\'i{}nez}}, \bibinfo {author} {\bibfnamefont {M.}~\bibnamefont
  {D\'i{}az}}, \bibinfo {author} {\bibfnamefont {E.}~\bibnamefont {Rom\'an}},
  \bibinfo {author} {\bibfnamefont {M.}~\bibnamefont {Ruano}}, \bibinfo
  {author} {\bibfnamefont {D.}~\bibnamefont {Llamosa~P.}}, \ and\ \bibinfo
  {author} {\bibfnamefont {Y.}~\bibnamefont {Huttel}},\ }\bibfield  {title}
  {\enquote {\bibinfo {title} {Generation of nanoparticles with adjustable size
  and controlled stoichiometry: Recent advances},}\ }\href {\doibase
  10.1021/la3022134} {\bibfield  {journal} {\bibinfo  {journal} {Langmuir}\
  }\textbf {\bibinfo {volume} {28}},\ \bibinfo {pages} {11241--11249} (\bibinfo
  {year} {2012})}\BibitemShut {NoStop}%
\bibitem [{\citenamefont {Frey}, \citenamefont {Corn},\ and\ \citenamefont
  {Weibel}(2006)}]{frey06}%
  \BibitemOpen
  \bibfield  {author} {\bibinfo {author} {\bibfnamefont {B.~L.}\ \bibnamefont
  {Frey}}, \bibinfo {author} {\bibfnamefont {R.~M.}\ \bibnamefont {Corn}}, \
  and\ \bibinfo {author} {\bibfnamefont {S.~C.}\ \bibnamefont {Weibel}},\
  }\enquote {\bibinfo {title} {Polarization-modulation approaches to
  reflection–absorption spectroscopy},}\ in\ \href {\doibase
  10.1002/0470027320.s2206} {\emph {\bibinfo {booktitle} {Handbook of
  Vibrational Spectroscopy}}}\ (\bibinfo  {publisher} {John Wiley {\&} Sons,
  Ltd},\ \bibinfo {year} {2006})\ pp.\ \bibinfo {pages}
  {1042--1056}\BibitemShut {NoStop}%
\bibitem [{\citenamefont {Boyer}\ \emph {et~al.}(2016)\citenamefont {Boyer},
  \citenamefont {Rivas}, \citenamefont {Tran}, \citenamefont {Verish},\ and\
  \citenamefont {Arumainayagam}}]{boyer16}%
  \BibitemOpen
  \bibfield  {author} {\bibinfo {author} {\bibfnamefont {M.~C.}\ \bibnamefont
  {Boyer}}, \bibinfo {author} {\bibfnamefont {N.}~\bibnamefont {Rivas}},
  \bibinfo {author} {\bibfnamefont {A.~A.}\ \bibnamefont {Tran}}, \bibinfo
  {author} {\bibfnamefont {C.~A.}\ \bibnamefont {Verish}}, \ and\ \bibinfo
  {author} {\bibfnamefont {C.~R.}\ \bibnamefont {Arumainayagam}},\ }\bibfield
  {title} {\enquote {\bibinfo {title} {The role of low-energy ($\leq$ 20ev)
  electrons in astrochemistry},}\ }\href {\doibase
  https://doi.org/10.1016/j.susc.2016.03.012} {\bibfield  {journal} {\bibinfo
  {journal} {Surface Science}\ }\textbf {\bibinfo {volume} {652}},\ \bibinfo
  {pages} {26 -- 32} (\bibinfo {year} {2016})}\BibitemShut {NoStop}%
\bibitem [{\citenamefont {Shulenberger}\ \emph {et~al.}(2019)\citenamefont
  {Shulenberger}, \citenamefont {Zhu}, \citenamefont {Tran}, \citenamefont
  {Abdullahi}, \citenamefont {Belvin}, \citenamefont {Lukens}, \citenamefont
  {Peeler}, \citenamefont {Mullikin}, \citenamefont {Cumberbatch},
  \citenamefont {Huang}, \citenamefont {Regovich}, \citenamefont {Zhou},
  \citenamefont {Heller}, \citenamefont {Markovic}, \citenamefont {Gates},
  \citenamefont {Buffo}, \citenamefont {Tano-Menka}, \citenamefont
  {Arumainayagam}, \citenamefont {Böhler}, \citenamefont {Swiderek},
  \citenamefont {Esmaili}, \citenamefont {Bass}, \citenamefont {Huels},\ and\
  \citenamefont {Sanche}}]{shulenberger19}%
  \BibitemOpen
  \bibfield  {author} {\bibinfo {author} {\bibfnamefont {K.~E.}\ \bibnamefont
  {Shulenberger}}, \bibinfo {author} {\bibfnamefont {J.~L.}\ \bibnamefont
  {Zhu}}, \bibinfo {author} {\bibfnamefont {K.}~\bibnamefont {Tran}}, \bibinfo
  {author} {\bibfnamefont {S.}~\bibnamefont {Abdullahi}}, \bibinfo {author}
  {\bibfnamefont {C.}~\bibnamefont {Belvin}}, \bibinfo {author} {\bibfnamefont
  {J.}~\bibnamefont {Lukens}}, \bibinfo {author} {\bibfnamefont
  {Z.}~\bibnamefont {Peeler}}, \bibinfo {author} {\bibfnamefont
  {E.}~\bibnamefont {Mullikin}}, \bibinfo {author} {\bibfnamefont {H.~M.}\
  \bibnamefont {Cumberbatch}}, \bibinfo {author} {\bibfnamefont
  {J.}~\bibnamefont {Huang}}, \bibinfo {author} {\bibfnamefont
  {K.}~\bibnamefont {Regovich}}, \bibinfo {author} {\bibfnamefont
  {A.}~\bibnamefont {Zhou}}, \bibinfo {author} {\bibfnamefont {L.}~\bibnamefont
  {Heller}}, \bibinfo {author} {\bibfnamefont {M.}~\bibnamefont {Markovic}},
  \bibinfo {author} {\bibfnamefont {L.}~\bibnamefont {Gates}}, \bibinfo
  {author} {\bibfnamefont {C.}~\bibnamefont {Buffo}}, \bibinfo {author}
  {\bibfnamefont {R.}~\bibnamefont {Tano-Menka}}, \bibinfo {author}
  {\bibfnamefont {C.~R.}\ \bibnamefont {Arumainayagam}}, \bibinfo {author}
  {\bibfnamefont {E.}~\bibnamefont {Böhler}}, \bibinfo {author} {\bibfnamefont
  {P.}~\bibnamefont {Swiderek}}, \bibinfo {author} {\bibfnamefont
  {S.}~\bibnamefont {Esmaili}}, \bibinfo {author} {\bibfnamefont {A.~D.}\
  \bibnamefont {Bass}}, \bibinfo {author} {\bibfnamefont {M.}~\bibnamefont
  {Huels}}, \ and\ \bibinfo {author} {\bibfnamefont {L.}~\bibnamefont
  {Sanche}},\ }\bibfield  {title} {\enquote {\bibinfo {title} {Electron-induced
  radiolysis of astrochemically relevant ammonia ices},}\ }\href {\doibase
  10.1021/acsearthspacechem.8b00169} {\bibfield  {journal} {\bibinfo  {journal}
  {ACS Earth and Space Chemistry}\ }\textbf {\bibinfo {volume} {3}},\ \bibinfo
  {pages} {800--810} (\bibinfo {year} {2019})}\BibitemShut {NoStop}%
\bibitem [{\citenamefont {{Jenniskens}}\ \emph {et~al.}(1993)\citenamefont
  {{Jenniskens}}, \citenamefont {{Baratta}}, \citenamefont {{Kouchi}},
  \citenamefont {{de Groot}}, \citenamefont {{Greenberg}},\ and\ \citenamefont
  {{Strazzulla}}}]{jenniskens93}%
  \BibitemOpen
  \bibfield  {author} {\bibinfo {author} {\bibfnamefont {P.}~\bibnamefont
  {{Jenniskens}}}, \bibinfo {author} {\bibfnamefont {G.~A.}\ \bibnamefont
  {{Baratta}}}, \bibinfo {author} {\bibfnamefont {A.}~\bibnamefont {{Kouchi}}},
  \bibinfo {author} {\bibfnamefont {M.~S.}\ \bibnamefont {{de Groot}}},
  \bibinfo {author} {\bibfnamefont {J.~M.}\ \bibnamefont {{Greenberg}}}, \ and\
  \bibinfo {author} {\bibfnamefont {G.}~\bibnamefont {{Strazzulla}}},\
  }\bibfield  {title} {\enquote {\bibinfo {title} {{Carbon dust formation on
  interstellar grains}},}\ }\href@noop {} {\bibfield  {journal} {\bibinfo
  {journal} {Astronomy {\&} Astrophysics}\ }\textbf {\bibinfo {volume} {273}},\
  \bibinfo {pages} {583} (\bibinfo {year} {1993})}\BibitemShut {NoStop}%
\bibitem [{\citenamefont {Chen}\ \emph {et~al.}(2014)\citenamefont {Chen},
  \citenamefont {Chuang}, \citenamefont {Caro}, \citenamefont {Nuevo},
  \citenamefont {Chu}, \citenamefont {Yih}, \citenamefont {Ip},\ and\
  \citenamefont {Wu}}]{chen14}%
  \BibitemOpen
  \bibfield  {author} {\bibinfo {author} {\bibfnamefont {Y.-J.}\ \bibnamefont
  {Chen}}, \bibinfo {author} {\bibfnamefont {K.-J.}\ \bibnamefont {Chuang}},
  \bibinfo {author} {\bibfnamefont {G.~M.~M.}\ \bibnamefont {Caro}}, \bibinfo
  {author} {\bibfnamefont {M.}~\bibnamefont {Nuevo}}, \bibinfo {author}
  {\bibfnamefont {C.-C.}\ \bibnamefont {Chu}}, \bibinfo {author} {\bibfnamefont
  {T.-S.}\ \bibnamefont {Yih}}, \bibinfo {author} {\bibfnamefont {W.-H.}\
  \bibnamefont {Ip}}, \ and\ \bibinfo {author} {\bibfnamefont {C.-Y.~R.}\
  \bibnamefont {Wu}},\ }\bibfield  {title} {\enquote {\bibinfo {title} {Vacuum
  ultraviolet emission spectrum measurement of a microwave-discharge
  hydrogen-flow lamp in several configurations: Application to photodesorption
  of {CO} ice},}\ }\href {\doibase 10.1088/0004-637X/781/1/15} {\bibfield
  {journal} {\bibinfo  {journal} {The Astrophysical Journal}\ }\textbf
  {\bibinfo {volume} {781}},\ \bibinfo {eid} {15} (\bibinfo {year}
  {2014})}\BibitemShut {NoStop}%
\bibitem [{\citenamefont {Baratta}, \citenamefont {Leto},\ and\ \citenamefont
  {Palumbo}(2002)}]{baratta02}%
  \BibitemOpen
  \bibfield  {author} {\bibinfo {author} {\bibfnamefont {G.~A.}\ \bibnamefont
  {Baratta}}, \bibinfo {author} {\bibfnamefont {G.}~\bibnamefont {Leto}}, \
  and\ \bibinfo {author} {\bibfnamefont {M.~E.}\ \bibnamefont {Palumbo}},\
  }\bibfield  {title} {\enquote {\bibinfo {title} {{A comparison of ion
  irradiation and UV photolysis of CH$_{4}$ and CH$_{3}$OH}},}\ }\href
  {\doibase 10.1051/0004-6361:20011835} {\bibfield  {journal} {\bibinfo
  {journal} {Astronomy {\&} Astrophysics}\ }\textbf {\bibinfo {volume} {384}},\
  \bibinfo {pages} {343--349} (\bibinfo {year} {2002})}\BibitemShut {NoStop}%
\bibitem [{\citenamefont {Fulvio}\ \emph {et~al.}(2014)\citenamefont {Fulvio},
  \citenamefont {Brieva}, \citenamefont {Cuylle}, \citenamefont {Linnartz},
  \citenamefont {Jäger},\ and\ \citenamefont {Henning}}]{fulvio14}%
  \BibitemOpen
  \bibfield  {author} {\bibinfo {author} {\bibfnamefont {D.}~\bibnamefont
  {Fulvio}}, \bibinfo {author} {\bibfnamefont {A.~C.}\ \bibnamefont {Brieva}},
  \bibinfo {author} {\bibfnamefont {S.~H.}\ \bibnamefont {Cuylle}}, \bibinfo
  {author} {\bibfnamefont {H.}~\bibnamefont {Linnartz}}, \bibinfo {author}
  {\bibfnamefont {C.}~\bibnamefont {Jäger}}, \ and\ \bibinfo {author}
  {\bibfnamefont {T.}~\bibnamefont {Henning}},\ }\bibfield  {title} {\enquote
  {\bibinfo {title} {A straightforward method for vacuum-ultraviolet flux
  measurements: The case of the hydrogen discharge lamp and implications for
  solid-phase actinometry},}\ }\href {\doibase 10.1063/1.4887067} {\bibfield
  {journal} {\bibinfo  {journal} {Applied Physics Letters}\ }\textbf {\bibinfo
  {volume} {105}},\ \bibinfo {pages} {014105} (\bibinfo {year}
  {2014})}\BibitemShut {NoStop}%
\bibitem [{\citenamefont {Teolis}, \citenamefont {Famá},\ and\ \citenamefont
  {Baragiola}(2007)}]{teolis07}%
  \BibitemOpen
  \bibfield  {author} {\bibinfo {author} {\bibfnamefont {B.~D.}\ \bibnamefont
  {Teolis}}, \bibinfo {author} {\bibfnamefont {M.}~\bibnamefont {Famá}}, \
  and\ \bibinfo {author} {\bibfnamefont {R.~A.}\ \bibnamefont {Baragiola}},\
  }\bibfield  {title} {\enquote {\bibinfo {title} {Low density solid ozone},}\
  }\href {\doibase 10.1063/1.2762215} {\bibfield  {journal} {\bibinfo
  {journal} {The Journal of Chemical Physics}\ }\textbf {\bibinfo {volume}
  {127}},\ \bibinfo {pages} {074507} (\bibinfo {year} {2007})}\BibitemShut
  {NoStop}%
\bibitem [{\citenamefont {Kuhn}, \citenamefont {Braslavsky},\ and\
  \citenamefont {Schmidt}(2004)}]{chemicalactinometry}%
  \BibitemOpen
  \bibfield  {author} {\bibinfo {author} {\bibfnamefont {H.~J.}\ \bibnamefont
  {Kuhn}}, \bibinfo {author} {\bibfnamefont {S.~E.}\ \bibnamefont
  {Braslavsky}}, \ and\ \bibinfo {author} {\bibfnamefont {R.}~\bibnamefont
  {Schmidt}},\ }\bibfield  {title} {\enquote {\bibinfo {title} {Chemical
  actinometry (iupac technical report)},}\ }\href {\doibase
  https://doi.org/10.1351/pac200476122105} {\bibfield  {journal} {\bibinfo
  {journal} {Pure and Applied Chemistry}\ }\textbf {\bibinfo {volume} {76}},\
  \bibinfo {pages} {2105 -- 2146} (\bibinfo {year} {2004})}\BibitemShut
  {NoStop}%
\bibitem [{\citenamefont {{Mart{\'\i}n-Dom{\'e}nech}}\ \emph
  {et~al.}(2015)\citenamefont {{Mart{\'\i}n-Dom{\'e}nech}}, \citenamefont
  {{Manzano-Santamar{\'\i}a}}, \citenamefont {{Mu{\~n}oz Caro}}, \citenamefont
  {{Cruz-D{\'\i}az}}, \citenamefont {{Chen}}, \citenamefont {{Herrero}},\ and\
  \citenamefont {{Tanarro}}}]{martindomenech15}%
  \BibitemOpen
  \bibfield  {author} {\bibinfo {author} {\bibfnamefont {R.}~\bibnamefont
  {{Mart{\'\i}n-Dom{\'e}nech}}}, \bibinfo {author} {\bibfnamefont
  {J.}~\bibnamefont {{Manzano-Santamar{\'\i}a}}}, \bibinfo {author}
  {\bibfnamefont {G.~M.}\ \bibnamefont {{Mu{\~n}oz Caro}}}, \bibinfo {author}
  {\bibfnamefont {G.~A.}\ \bibnamefont {{Cruz-D{\'\i}az}}}, \bibinfo {author}
  {\bibfnamefont {Y.~J.}\ \bibnamefont {{Chen}}}, \bibinfo {author}
  {\bibfnamefont {V.~J.}\ \bibnamefont {{Herrero}}}, \ and\ \bibinfo {author}
  {\bibfnamefont {I.}~\bibnamefont {{Tanarro}}},\ }\bibfield  {title} {\enquote
  {\bibinfo {title} {{UV photoprocessing of CO$_{2}$ ice: a complete
  quantification of photochemistry and photon-induced desorption processes}},}\
  }\href {\doibase 10.1051/0004-6361/201526003} {\bibfield  {journal} {\bibinfo
   {journal} {Astronomy {\&} Astrophysics}\ }\textbf {\bibinfo {volume}
  {584}},\ \bibinfo {eid} {A14} (\bibinfo {year} {2015})}\BibitemShut {NoStop}%
\bibitem [{\citenamefont {Cottin}, \citenamefont {Moore},\ and\ \citenamefont
  {Benilan}(2003)}]{cottin03}%
  \BibitemOpen
  \bibfield  {author} {\bibinfo {author} {\bibfnamefont {H.}~\bibnamefont
  {Cottin}}, \bibinfo {author} {\bibfnamefont {M.~H.}\ \bibnamefont {Moore}}, \
  and\ \bibinfo {author} {\bibfnamefont {Y.}~\bibnamefont {Benilan}},\
  }\bibfield  {title} {\enquote {\bibinfo {title} {Photodestruction of relevant
  interstellar molecules in ice mixtures},}\ }\href {\doibase 10.1086/375149}
  {\bibfield  {journal} {\bibinfo  {journal} {The Astrophysical Journal}\
  }\textbf {\bibinfo {volume} {590}},\ \bibinfo {pages} {874--881} (\bibinfo
  {year} {2003})}\BibitemShut {NoStop}%
\bibitem [{\citenamefont {{Gerakines}}, \citenamefont {{Schutte}},\ and\
  \citenamefont {{Ehrenfreund}}(1996)}]{gerakines96}%
  \BibitemOpen
  \bibfield  {author} {\bibinfo {author} {\bibfnamefont {P.~A.}\ \bibnamefont
  {{Gerakines}}}, \bibinfo {author} {\bibfnamefont {W.~A.}\ \bibnamefont
  {{Schutte}}}, \ and\ \bibinfo {author} {\bibfnamefont {P.}~\bibnamefont
  {{Ehrenfreund}}},\ }\bibfield  {title} {\enquote {\bibinfo {title}
  {{Ultraviolet processing of interstellar ice analogs. I. Pure ices.}}}\
  }\href@noop {} {\bibfield  {journal} {\bibinfo  {journal} {Astronomy {\&}
  Astrophysics}\ }\textbf {\bibinfo {volume} {312}},\ \bibinfo {pages}
  {289--305} (\bibinfo {year} {1996})}\BibitemShut {NoStop}%
\bibitem [{\citenamefont {Accolla}\ \emph {et~al.}(2011)\citenamefont
  {Accolla}, \citenamefont {Congiu}, \citenamefont {Dulieu}, \citenamefont
  {Manico}, \citenamefont {Chaabouni}, \citenamefont {Matar}, \citenamefont
  {Mokrane}, \citenamefont {Lemaire},\ and\ \citenamefont
  {Pirronello}}]{accolla11}%
  \BibitemOpen
  \bibfield  {author} {\bibinfo {author} {\bibfnamefont {M.}~\bibnamefont
  {Accolla}}, \bibinfo {author} {\bibfnamefont {E.}~\bibnamefont {Congiu}},
  \bibinfo {author} {\bibfnamefont {F.}~\bibnamefont {Dulieu}}, \bibinfo
  {author} {\bibfnamefont {G.}~\bibnamefont {Manico}}, \bibinfo {author}
  {\bibfnamefont {H.}~\bibnamefont {Chaabouni}}, \bibinfo {author}
  {\bibfnamefont {E.}~\bibnamefont {Matar}}, \bibinfo {author} {\bibfnamefont
  {H.}~\bibnamefont {Mokrane}}, \bibinfo {author} {\bibfnamefont {J.~L.}\
  \bibnamefont {Lemaire}}, \ and\ \bibinfo {author} {\bibfnamefont
  {V.}~\bibnamefont {Pirronello}},\ }\bibfield  {title} {\enquote {\bibinfo
  {title} {{Changes in the morphology of interstellar ice analogues after
  hydrogen atom exposure}},}\ }\href {\doibase 10.1039/c0cp01462a} {\bibfield
  {journal} {\bibinfo  {journal} {Physical Chemistry Chemical Physics}\
  }\textbf {\bibinfo {volume} {13}},\ \bibinfo {pages} {8037--8045} (\bibinfo
  {year} {2011})}\BibitemShut {NoStop}%
\bibitem [{\citenamefont {Snyder}, \citenamefont {Hsu},\ and\ \citenamefont
  {Krimm}(1978)}]{snyder78}%
  \BibitemOpen
  \bibfield  {author} {\bibinfo {author} {\bibfnamefont {R.}~\bibnamefont
  {Snyder}}, \bibinfo {author} {\bibfnamefont {S.}~\bibnamefont {Hsu}}, \ and\
  \bibinfo {author} {\bibfnamefont {S.}~\bibnamefont {Krimm}},\ }\bibfield
  {title} {\enquote {\bibinfo {title} {Vibrational spectra in the {C-H}
  stretching region and the structure of the polymethylene chain},}\ }\href
  {\doibase https://doi.org/10.1016/0584-8539(78)80167-6} {\bibfield  {journal}
  {\bibinfo  {journal} {Spectrochimica Acta Part A: Molecular Spectroscopy}\
  }\textbf {\bibinfo {volume} {34}},\ \bibinfo {pages} {395 -- 406} (\bibinfo
  {year} {1978})}\BibitemShut {NoStop}%
\bibitem [{\citenamefont {Snyder}\ and\ \citenamefont
  {Schachtschneider}(1963)}]{snyder63}%
  \BibitemOpen
  \bibfield  {author} {\bibinfo {author} {\bibfnamefont {R.}~\bibnamefont
  {Snyder}}\ and\ \bibinfo {author} {\bibfnamefont {J.}~\bibnamefont
  {Schachtschneider}},\ }\bibfield  {title} {\enquote {\bibinfo {title}
  {Vibrational analysis of the n-paraffins—i: Assignments of infrared bands
  in the spectra of {C$_3$H$_8$} through {n-C$_{19}$H$_{40}$}},}\ }\href
  {\doibase https://doi.org/10.1016/0371-1951(63)80095-8} {\bibfield  {journal}
  {\bibinfo  {journal} {Spectrochimica Acta}\ }\textbf {\bibinfo {volume}
  {19}},\ \bibinfo {pages} {85 -- 116} (\bibinfo {year} {1963})}\BibitemShut
  {NoStop}%
\bibitem [{\citenamefont {{Socrates}}(2001)}]{socrates}%
  \BibitemOpen
  \bibfield  {author} {\bibinfo {author} {\bibfnamefont {G.}~\bibnamefont
  {{Socrates}}},\ }\href@noop {} {\emph {\bibinfo {title} {Infrared and Raman
  characteristic group frequencies}}},\ edited by\ \bibinfo {editor}
  {\bibfnamefont {J.~W. .~S.}\ \bibnamefont {Ltd}}\ (\bibinfo  {publisher}
  {John Wiley \& Sons},\ \bibinfo {year} {2001})\BibitemShut {NoStop}%
\bibitem [{\citenamefont {McNesby}\ and\ \citenamefont
  {Okabe}(2007)}]{vuv_photochemistry}%
  \BibitemOpen
  \bibfield  {author} {\bibinfo {author} {\bibfnamefont {J.~R.}\ \bibnamefont
  {McNesby}}\ and\ \bibinfo {author} {\bibfnamefont {H.}~\bibnamefont
  {Okabe}},\ }\enquote {\bibinfo {title} {Vacuum ultraviolet photochemistry},}\
  in\ \href {\doibase 10.1002/9780470133330.ch3} {\emph {\bibinfo {booktitle}
  {Advances in Photochemistry}}}\ (\bibinfo  {publisher} {John Wiley \& Sons,
  Ltd},\ \bibinfo {year} {2007})\ pp.\ \bibinfo {pages} {157--240}\BibitemShut
  {NoStop}%
\bibitem [{\citenamefont {{NIST Mass Spectrometry Data Center}}(2020)}]{nist}%
  \BibitemOpen
  \bibfield  {author} {\bibinfo {author} {\bibnamefont {{NIST Mass Spectrometry
  Data Center}}},\ }\enquote {\bibinfo {title} {{NIST} chemistry webbook,
  {NIST} standard reference database number 69},}\ \ (\bibinfo  {publisher}
  {National Institute of Standards and Technology},\ \bibinfo {year} {2020})\
  Chap.\ \bibinfo {chapter} {Mass Spectra}\BibitemShut {NoStop}%
\end{thebibliography}%

\end{document}